\RequirePackage{fix-cm}
\documentclass[smallextended]{svjour3}       
\smartqed  

\usepackage{bm,amsmath,cite}  
\usepackage{amsfonts} 
\usepackage{graphicx} 

\usepackage[english]{babel}
\usepackage{siunitx,bm}
\usepackage{graphicx,rotating,multirow}
\usepackage{bm,amsmath,amssymb,amsfonts,color,mathrsfs,cite,cuted} 
\usepackage{relsize,exscale}

\journalname{Eur. Phys. J. Plus}

\newcommand{\AM}{ Annals Math. }
\newcommand{\AnM}{ Annals Math. }

\newcommand{\APB}{ Ann. Phys. (Berlin) }
\newcommand{\APNY}{ Ann. Phys. (N.Y.) }

\newcommand{\CMP}{ Commun. Math. Phys. }

\newcommand{\CPL}{ Chem. Phys. Lett. }
\newcommand{\CRA}{ C. R. Acad. Sci. Ser. A }
\newcommand{\EJP}{ Eur. J. Phys. }

\newcommand{\EPJP}{ Eur. Phys. J. Plus }

\newcommand{\IEEETMT}{ IEEE T. Microw. Theory }

\newcommand{\IJQC}{ Int. J. Quantum Chem. }
\newcommand{\JAP}{ J. Appl. Phys. }

\newcommand{\JFA}{ J. Funct. Anal. }

\newcommand{\JMAA}{ J. Math. Anal. Appl. }

\newcommand{\JMPC}{ J. Math. Phys. Camb. }
\newcommand{\jpa}{ J. Phys. A }

\newcommand{\JSP}{ J. Stat. Phys. }

\newcommand{\MPC}{ Math. Proc. Cambridge }
\newcommand{\NJP}{ New J. Phys. }

\newcommand{\PA}{ Physica A }

\newcommand{\PJP}{ Pramana - J. Phys. }
\newcommand{\PLA}{ Phys. Lett. A }
\newcommand{\PR}{ Phys. Rev. }
\newcommand{\PRA}{ Phys. Rev. A }
\newcommand{\PRB}{ Phys. Rev. B }

\newcommand{\PRE}{ Phys. Rev. E }
\newcommand{\PRL}{ Phys. Rev. Lett. }
\newcommand{\PRe}{ Phys. Rep. }
\newcommand{\PRXQ}{ PRX Quantum }

\newcommand{\RMP}{ Rev. Mod. Phys. }

\newcommand{\SR}{ Sci. Rep. - UK }

\begin{document}

\title{Quantum-information theory of a Dirichlet ring with Aharonov-Bohm field\thanks{Research was supported by Competitive Research Project No. 2002143087 from the Research Funding Department, Vice Chancellor for Research and Graduate Studies, University of Sharjah.}
}

\titlerunning{Quantum-information theory of the Aharonov-Bohm ring}        

\author{O. Olendski}

\institute{O. Olendski \at
              Department of Applied Physics and Astronomy, University of Sharjah, P.O. Box 27272, Sharjah, United Arab Emirates \\
              \email{oolendski@sharjah.ac.ae}           
}

\date{Received: date / Accepted: date}

\maketitle

\begin{abstract}
Shannon quantum information entropies $S_{\rho,\gamma}(\phi_{AB},r_0)$, Fisher informations $I_{\rho,\gamma}(\phi_{AB},r_0)$, Onicescu energies $O_{\rho,\gamma}(\phi_{AB},r_0)$ and R\'{e}nyi entropies $R_{\rho,\gamma}(\phi_{AB},r_0;\alpha)$ are calculated both in the position (subscript $\rho$) and momentum ($\gamma$) spaces as functions of the inner radius $r_0$ for the two-dimensional Dirichlet unit-width annulus threaded by the Aharonov-Bohm (AB) flux $\phi_{AB}$. Small (huge) values of $r_0$ correspond to the thick (thin) rings with extreme of $r_0=0$ describing the dot. Discussion is based on the analysis of the corresponding position $\Psi_{nm}(\phi_{AB},r_0;{\bf r})$ and momentum $\Phi_{nm}(\phi_{AB},r_0;{\bf k})$ waveforms, with $n$ and $m$ being principal and magnetic quantum indices, respectively: the former allows an analytic expression at any AB field whereas for the latter it is true at the flux-free configuration, $\phi_{AB}=0$, only. It is shown, in particular, that the position Shannon entropy $S_{\rho_{nm}}(\phi_{AB},r_0)$ [Onicescu energy $O_{\rho_{nm}}(\phi_{AB},r_0)$] grows logarithmically [decreases as $1/r_0$] with large $r_0$ tending to the same asymptote $S_\rho^{asym}=\ln(4\pi r_0)-1$ [$O_\rho^{asym}=3/(4\pi r_0)$] for all orbitals whereas their Fisher counterpart $I_{\rho_{nm}}(\phi_{AB},r_0$) approaches in the same regime the $m$-independent  limit mimicking in this way the energy spectrum variation with $r_0$, which for the thin structures exhibits quadratic dependence on the principal index. Frequency of the fading oscillations of the radial parts of the wave vector functions $\Phi_{nm}(\phi_{AB},r_0;{\bf k})$ increases with the inner radius what results in the identical $r_0\gg1$ asymptote for all momentum Shannon entropies $S_{\gamma_{nm}}(\phi_{AB};r_0)$ with the alike $n$ and different $m$. The same limit causes the Fisher momentum components $I_\gamma(\phi_{AB},r_0)$ to grow exponentially with $r_0$. Based on these calculations, properties of the complexities $e^SO$ are addressed too. Among many findings on the R\'{e}nyi entropy, it is proved that the lower limit $\alpha_{TH}$ of the semi-infinite range of the dimensionless coefficient $\alpha$, where the momentum component of this one-parameter entropy exists, is \textit{not} influenced by the radius; in particular, the change of the topology from the simply, $r_0=0$, to the doubly, $r_0>0$, connected domain is \textit{un}able to change $\alpha_{TH}=2/5$. AB field influence on the measures is calculated too. Parallels are drawn to the geometry with volcano-shape confining potential and similarities and differences between them are discussed.
\end{abstract}

\section{Introduction}\label{sec_Intro}
To describe the processes occurring in nanoscale structures, quantum-information theory operates with some functionals of the position $\rho({\bf r})$ and wave vector (or momentum ${\bf p}=\hbar{\bf k}$) $\gamma({\bf k})$ densities which are modulo squares of the corresponding one-particle wave functions $\Psi({\bf r})$ and $\Phi({\bf k})$:
\begin{subequations}\label{Densities1}
\begin{align}
\label{DensityX1}
\rho_\mathtt{n}({\bf r})&=\left|\Psi_\mathtt{n}({\bf r})\right|^2\\
\label{DensityP1}
\gamma_\mathtt{n}({\bf k})&=\left|\Phi_\mathtt{n}({\bf k})\right|^2,
\end{align}
\end{subequations}
with $\mathtt{n}=0,1,\ldots$ being a combined quantum number of the corresponding orbital in the $\mathtt{d}$-dimensional ($\mathtt{d}$D) space. $\Psi_\mathtt{n}({\bf r})$ and $\Phi_\mathtt{n}({\bf k})$ are Fourier transforms of each other
\begin{subequations}\label{Fourier1}
\begin{align}\label{Fourier1_1}
\Phi_\mathtt{n}({\bf k})&=\frac{1}{(2\pi)^{\mathtt{d}/2}}\int_{\mathcal{D}_\rho^{(\mathtt{d})}}\Psi_\mathtt{n}({\bf r})e^{-i{\bf kr}}d{\bf r},\\
\label{Fourier1_2}
\Psi_\mathtt{n}({\bf r})&=\frac{1}{(2\pi)^{\mathtt{d}/2}}\int_{\mathcal{D}_\gamma^{(\mathtt{d})}}\Phi_\mathtt{n}({\bf k})e^{i{\bf rk}}d{\bf k},
\end{align}
\end{subequations}
with $\mathcal{D}_\rho^{(\mathtt{d})}$ and $\mathcal{D}_\gamma^{(\mathtt{d})}$  being the domains where these dependencies are defined. Each of the sets is orthonormalized:
\begin{equation}\label{OrthoNormality1}
\int_{\mathcal{D}_\rho^{(\mathtt{d})}}\Psi_\mathtt{n'}^\ast({\bf r})\Psi_\mathtt{n}d{\bf r}=\int_{\mathcal{D}_\gamma^{(\mathtt{d})}}\Phi_\mathtt{n'}^*({\bf k})\Phi_\mathtt{n}({\bf k})d{\bf k}=\delta_\mathtt{n'n},
\end{equation}
$\delta_\mathtt{n'n}$ is a Kronecker delta, and position component is a solution of the one-particle $\mathtt{d}$D Schr\"{o}dinger equation
\begin{equation}\label{Schrodinger1}
\widehat{H}\Psi_\mathtt{n}({\bf r})=E_\mathtt{n}\Psi_\mathtt{n}({\bf r})
\end{equation}
with the Hamiltonian
\begin{equation}\label{Hamiltonian1}
\widehat{H}=\frac{1}{2M}\left[i\hbar{\bm\nabla}_{\bf r}+q{\bf A}({\bf r})\right]^2+V({\bf r}),
\end{equation}
where $V({\bf r})$ and ${\bf A}({\bf r})$ are the external electrostatic and vector potentials, respectively, in which the particle with  mass $M$, charge $q$ and energy $E_\mathtt{n}$ is moving. Magnetic field ${\bf B}({\bf r})$ enters the above equation via the vector potential, ${\bf B}({\bf r})={\bm\nabla}_{\bf r}\times{\bf A}({\bf r})$.

Probably, the most famous and frequently used functional is Shannon quantum-information entropy \cite{Shannon1,Shannon2}  whose expressions in the position $S_\rho$ and wave vector $S_\gamma$ representations are:
\begin{subequations}\label{Shannon1}
\begin{align}\label{Shannon1_R}
S_{\rho_\mathtt{n}}&=-\int_{\mathcal{D}_\rho^{(\mathtt{d})}}\rho_\mathtt{n}({\bf r})\ln\rho_\mathtt{n}({\bf r})d{\bf r}\\
\label{Shannon1_K}
S_{\gamma_\mathtt{n}}&=-\int_{\mathcal{D}_\gamma^{(\mathtt{d})}}\gamma_\mathtt{n}({\bf k})\ln\gamma_\mathtt{n}({\bf k})d{\bf k}.
\end{align}
\end{subequations} 
They quantify the amount of \textit{un}available information about the properties in the corresponding space: 
larger (smaller) values of $S_\rho$ or $S_\gamma$ mean less (more) knowledge about the particle location or momentum, respectively. Expressions~\eqref{Shannon1} extend to the continuous case the Shannon entropy for the $N$ discrete events with probabilities $p_i$, $i=1,2,\ldots,N$ \cite{Shannon1}:
\begin{equation}\label{ShannonDiscrete1}
S_p=-\sum_{i=1}^Np_i\ln p_i,
\end{equation}
with $N$ being finite or infinite, and $\sum_{i=1}^Np_i=1$. Since the discrete probabilities can not be greater than unity, $0\leq p_i\leq1$, expression from Eq.~\eqref{ShannonDiscrete1} is always positive. For their part, functionals from Eqs.~\eqref{Shannon1} might turn negative \cite{Shannon2}. Furthermore, the two components for the same orbital are not independent from each other but obey the fundamental relation establishing the limit of our total knowledge about the corpuscle behavior \cite{Bialynicki1,Beckner1}:
\begin{equation}\label{ShannonInequality1}
S_{t_\mathtt{n}}\geq\mathtt{d}(1+\ln\pi).
\end{equation}
with $S_t$ being the sum of the two entropies:
\begin{equation}\label{ShannonSum1}
S_t=S_\rho+S_\gamma.
\end{equation}
It is known that fundamental inequality~\eqref{ShannonInequality1} presents a much more general base for defining  'uncertainty' \cite{Coles1} than the famous Heisenberg relation since the former always holds true whereas the latter is either violated or meaningless, in particular, for the structures with non-Dirichlet boundary conditions \cite{Bialynicki3,Olendski10,Olendski11,Olendski5}. Besides this fundamental relation, Shannon entropy is indispensable in many other fields of nano physics; for example, its evaluation on eigenfunctions of quantum systems has been investigated in connection with a method for the approximate description of pure states based on the maximum entropy principle \cite{Plastino1}. Let us also mention that an important characteristics of the Shannon measure is its additivity \cite{Shannon1}:
\begin{equation}\label{ShannonAdditivity1}
S_{fg}=S_f+S_g,
\end{equation}
meaning that the uncertainty of one event $f$ is not influenced by the second set $g$, if $f$ and $g$ are the two independent happenings; in other words, total information acquired from the independent distributions is the sum of its counterparts for each of them.

Contrary to the Shannon functionals, which are global identifiers of the system, another quantum-information measure introduced by statistician and biologist R. A. Fisher almost simultaneously with the birth of quantum mechanics \cite{Fisher1,Frieden1}, is a local descriptor of uncertainty:
\begin{subequations}\label{Fisher1}
\begin{align}\label{Fisher1_R}
I_{\rho_\mathtt{n}}&=\int_{\mathcal{D}_\rho^{(\mathtt{d})}}\rho_\mathtt{n}({\bf r})\left|{\bm\nabla}_{\bf r}\ln\rho_\mathtt{n}({\bf r})\right|^2\!\!d{\bf r}=\int_{\mathcal{D}_\rho^{(\mathtt{d})}}\frac{\left|{\bm\nabla}_{\bf r}\rho_\mathtt{n}({\bf r})\right|^2}{\rho_\mathtt{n}({\bf r})}d{\bf r}\\
\label{Fisher1_K}
I_{\gamma_\mathtt{n}}&=\int_{\mathcal{D}_\gamma^{(\mathtt{d})}}\gamma_\mathtt{n}({\bf k})\left|{\bm\nabla}_{\bf k}\ln\gamma_\mathtt{n}({\bf k})\right|^2\!\!d{\bf k}=\int_{\mathcal{D}_\gamma^{(\mathtt{d})}}\frac{\left|{\bm\nabla}_{\bf k}\gamma_\mathtt{n}({\bf k})\right|^2}{\gamma_\mathtt{n}({\bf k})}\,d{\bf k}.
\end{align}
\end{subequations}
Due to the presence of the gradients, these integrals quantify the speed of change of the probability function in the corresponding domain. In one of its most important applications, the position Fisher information defines the kinetic energy of the many-particle system, which allows to reformulate the quantum mechanical variation principle as a principle of minimal information \cite{Sears1}. In a closely related development, since the Lagrangian of the physical phenomena described by the Schr\"{o}dinger equation contains a squared gradient term, the latter can be obtained from the variational principle based on Fisher information \cite{Frieden1}. The product of the position and momentum components is a measure of the overall intensity of variation in both spaces. Contrary to Shannon uncertainty law, Eq.~\eqref{ShannonInequality1}, similar universal relation for  $I_\rho I_\gamma$ is not known \cite{Plastino2} though for several particular systems some inequalities have been derived \cite{Stam1,Dembo1,Romera2,Dehesa1,SannchezMoreno1,SannchezMoreno2}. Beyond physics, the functional $I$ is being actively used in, e.g., economics, analysis of cancer growth, transport processes, etc. \cite{Frieden1}. Let us also point out that originally this functional was introduced as a way of measuring the amount of information that an observable random variable $\cal X$ carries about an unknown parameter $\Theta$ upon which the probability $f({\cal X}|\Theta)$ depends, with $f({\cal X}|\Theta)$ being the probability density function for $\cal X$ conditioned on the value of $\Theta$. In this case, the Fisher information is defined by an expression involving the partial derivative of $\ln\!f$ with respect to the parameter $\Theta$ \cite{Fisher1}:
\begin{equation}\label{FisherOriginal1}
I_{\cal X}(\Theta)=\int_{\Omega_{{\cal X}|\Theta}}\!\!f({\cal X}|\Theta)\left[\frac{\partial\ln\!f({\cal X}|\Theta)}{\partial\Theta}\right]^2d{\cal X},
\end{equation}
with $\Omega_{{\cal X}|\Theta}$ being a set of all admissible values of $\cal X$. Instead, Eqs.~\eqref{Fisher1} involve the partial derivatives of the logarithm of the densities $\rho$ and $\gamma$ with respect to the spatial and wave vector coordinates, respectively, what means that our treatment is related to a particular case of Fisher information in which the parameter $\Theta$ corresponds to, say, the spatial shifts of the density $\rho$.

Deviations from the uniform distributions are described by Onicescu energies or disequilibria:
\begin{subequations}\label{Onicescu1}
\begin{align}\label{Onicescu1_R}
O_{\rho_\mathtt{n}}&=\int_{\mathcal{D}_\rho^{(\mathtt{d})}}\rho_\mathtt{n}^2({\bf r})d{\bf r}\\
\label{Onicescu1_K}
O_{\gamma_\mathtt{n}}&=\int_{\mathcal{D}_\gamma^{(\mathtt{d})}}\gamma_\mathtt{n}^2({\bf k})d{\bf k}.
\intertext{This quantity was originally introduced by the Romanian mathematician for the discrete distribution  \cite{Onicescu1}:}
\label{Onicescu1_P}
O_p&=\sum_{i=1}^Np_i^2,
\end{align}
\end{subequations}
where it was called 'information energy'. As quite often happens in scientific terminology, the word 'energy' in the context of Eqs.~\eqref{Onicescu1} is obscure and confusing since these measures depend on
the probability densities only and do not involve the system’s Hamiltonian. Multiplying position and momentum components, one gets a gross estimation of the deflection of the state from homogeneity with conjecture \cite{Ghosal1}:
\begin{equation}\label{OnicescuInequality1}
O_\rho O_\gamma\leq\frac{1}{(2\pi)^\mathtt{d}}.
\end{equation}

One-parameter generalization of the Shannon entropy that preserves its additivity,
\begin{equation}\label{RenyiAdditivity1}
R_{fg}(\alpha)=R_f(\alpha)+R_g(\alpha),
\end{equation}
was introduced by A. R\'{e}nyi \cite{Renyi1,Renyi2}:
\begin{equation}\label{RenyiDiscrete1}
R_p(\alpha)=\frac{1}{1-\alpha}\sum_{i=1}^N\ln p_i^\alpha
\end{equation}
what for the continuous dependencies transforms to
\begin{subequations}\label{Renyi1}
\begin{align}\label{Renyi1_R}
R_{\rho_\mathtt{n}}(\alpha)&=\frac{1}{1-\alpha}\ln\!\left(\int_{\mathcal{D}_\rho^{(\mathtt{d})}}\rho_\mathtt{n}^\alpha({\bf r})d{\bf r}\right)\\
\label{Renyi1_K}
R_{\gamma_\mathtt{n}}(\alpha)&=\frac{1}{1-\alpha}\ln\!\left(\!\int_{\mathcal{D}_\gamma^{(\mathtt{d})}}\gamma_\mathtt{n}^\alpha({\bf k})d{\bf k}\right).
\end{align}
\end{subequations}
In the limit of the unit R\'{e}nyi parameter, $\alpha\rightarrow1$, the measures from Eqs.~\eqref{RenyiDiscrete1} and \eqref{Renyi1} turn, according to  the l'H\^{o}pital’s rule, into their Shannon counterparts from Eqs.~\eqref{ShannonDiscrete1} and \eqref{Shannon1}, respectively:
\begin{equation}\label{RenyiToShannon1}
R(1)=S.
\end{equation}
The same holds true for the R\'{e}nyi uncertainty relation \cite{Bialynicki2,Zozor1}:
\begin{equation}\label{RenyiUncertainty1}
R_{\rho_\mathtt{n}}(\alpha)+R_{\gamma_\mathtt{n}}(\beta)\geq-\frac{\mathtt{d}}{2}\left(\frac{1}{1-\alpha}\ln\frac{\alpha}{\pi}+\frac{1}{1-\beta}\ln\frac{\beta}{\pi}\right),
\end{equation}
with the positive parameters $\alpha$ and $\beta$ being conjugated as
\begin{equation}\label{RenyiUncertainty2}
\frac{1}{\alpha}+\frac{1}{\beta}=2;
\end{equation}
namely, the limit $\alpha\rightarrow1$ transforms Eq.~\eqref{RenyiUncertainty1} into its Shannon counterpart, Eq.~\eqref{ShannonInequality1}. Uncertainty relations find their indispensable applications in data compression, quantum cryptography, entanglement witnessing, quantum metrology and other tasks employing correlations between the position and momentum components of the information measures \cite{Wehner1,Jizba2,Coles1,Toscano1,Hertz1,Wang1}. Let us also note that disequilibrium is a particular case of the R\'{e}nyi entropy too:
\begin{equation}\label{OnicescuToRenyi1}
O=e^{-R(2)}.
\end{equation}

Another one-parameter generalization of the Shannon measure is celebrated Tsallis entropy \cite{Tsallis1} that for the discrete and continuous cases is written as
\begin{subequations}\label{Tsallis1}
\begin{align}\label{Tsallis1_D}
T_p(\alpha)&=\frac{1}{\alpha-1}\sum_{i=1}^N(1-p_i^\alpha)\\
\label{Tsallis1_R}
T_{\rho_\mathtt{n}}(\alpha)&=\frac{1}{\alpha-1}\left(1-\int_{\mathcal{D}_\rho^{(\mathtt{d})}}\rho_\mathtt{n}^\alpha({\bf r})d{\bf r}\right)\\
\label{Tsallis1_K}
T_{\gamma_\mathtt{n}}(\alpha)&=\frac{1}{\alpha-1}\left(1-\int_{\mathcal{D}_\gamma^{(\mathtt{d})}}\gamma_\mathtt{n}^\alpha({\bf k})d{\bf k}\right).
\end{align}
\end{subequations}
Similar to the R\'{e}nyi measure, at $\alpha\rightarrow1$ l'H\^{o}pital's rule degenerates it to the Shannon dependence too:
\begin{equation}\label{TsallisToShannon1}
T(1)=S.
\end{equation}
Tsallis entropy is only \textit{pseudo}-additive:
\begin{equation}\label{AdditivityTsallis1}
T_{fg}(\alpha)=T_f(\alpha)+T_g(\alpha)+(1-\alpha)T_f(\alpha)T_{g}(\alpha).
\end{equation}
Eqs.~\eqref{Tsallis1_R} and~\eqref{Tsallis1_K} manifest that Tsallis entropy for the continuous case has a dimensionality problem; however, both sides of the corresponding uncertainty relation \cite{Rajagopal1}
\begin{equation}\label{TsallisInequality1}
\left(\frac{\alpha}{\pi}\right)^{\mathtt{d}/(4\alpha)}\!\!\left[1+(1-\alpha)T_{\rho_\mathtt{n}}(\alpha)\right]^{1/(2\alpha)}\!\!\geq\!\!\left(\frac{\beta}{\pi}\right)^{\mathtt{d}/(4\beta)}\!\!\left[1+(1-\beta)T_{\gamma_\mathtt{n}}(\beta)\right]^{1/(2\beta)},
\end{equation}
which is valid inside the interval $1/2\leq\alpha\leq1$, are measured in the same units and can be compared to each other; namely, Eq.~\eqref{TsallisInequality1} at $\alpha=1$ degenerates into the trivial identity and close to it takes the form \cite{Olendski3}:
\begin{equation}\label{TsallisInequality1_1}
\frac{1+\left[-2S_\rho+\mathtt{d}(1+\ln\pi)\right](\alpha-1)/4}{\pi^{\mathtt{d}/4}}\geq\frac{1+\left[2S_\gamma-\mathtt{d}(1+\ln\pi)\right](\alpha-1)/4}{\pi^{\mathtt{d}/4}},\quad\alpha\rightarrow1.
\end{equation}
At the opposite edge of one half, the lowest energy orbital saturates it too whereas for any excited level it continues to be a strict inequality. Since all these properties for the QR are very similar to the previous geometries\cite{Olendski1,Olendski3,Olendski6,Olendski7,Olendski8}, they will not be discussed here.

Based on these measures, miscellaneous complexities are built, such as, for example, combination of Shannon entropy and disequilibrium,
\begin{subequations}\label{Complexities1}
\begin{align}\label{CGL1}
CGL=e^SO
\intertext{with its properties discussed first by R. G. Catal\'{a}n, J. Garay,  and R. L\'{o}pez-Ruiz \cite{Catalan1}; Shannon-Fisher product  \cite{Dembo1,Vignat1}}
\label{ShannonFisher1}
\frac{1}{2\pi e}\,e^{2S/\mathtt{d}}I;
\intertext{interplay of the R\'{e}nyi entropy and Onicescu energy \cite{Yamano1,Antolin1}}
\label{RenyiOnicescu1}
e^{R(\alpha)}O;
\intertext{Fisher-R\'{e}nyi product \cite{Antolin1,Romera1}}
\label{FisherRenyi1}
\frac{1}{2\pi e}\,e^{2R(\alpha)/\mathtt{d}}I,
\end{align}
\end{subequations}
and others, see References \cite{Sen1,Toranzo1} and literature therein. Below, for our system we will analyze the $CGL$ measure that tries to incorporate in one number both the randomness represented by the first multiplier in the right-hand side of Eq.~\eqref{CGL1} as well as the deviation of the probability density from homogeneity described by the second term there. This complexity in either $\mathtt{d}$D space can not be smaller than unity, $CGL\geq1$ \cite{LopezRosa1}.

All mentioned above quantities are under the vigorous theoretical analysis for different nanostructures that is additionally spurred by the experimental advances in evaluating them; e.g., Shannon \cite{Lukin1,Niknam1} and R\'{e}nyi \cite{Niknam1,Islam1,Kaufman1,Brydges1} entropies were detected and estimated for several many-body systems. In a very recent development, Shannon, R\'{e}nyi and Tsallis entropies together with the Fisher informations and disequilibria were calculated for the circular quantum dot (QD) with the Dirichlet or Neumann boundary conditions (BCs) and their similarities and differences were pointed out \cite{Olendski1}. Present endeavor extends this research to the case of the quantum ring (QR), i.e., a 2D planar domain where the charged particle moves inside the annular region of the width $a$ with the inner radius  $r_0$ that is additionally pierced at its center by the Aharonov-Bohm (AB) field \cite{Aharonov1}. Here, we address the Dirichlet BC with the position waveform vanishing at the ring edges, $\left.\Psi({\bf r})\right|_{\cal S}=0$, whereas a discussion of the Neumann requirement when the normal derivative turns to zero at the confining surfaces $\cal S$,
%
$$
\left.\frac{\partial\Psi({\bf r})}{\partial{\bf n}}\right|_{\cal S}=0,
$$
%
${\bf n}$ is a unit normal to $\cal S$, is deferred to the separate presentation. Different BCs are relevant for the correct theoretical description of the miscellaneous materials which might be used in design of the quantum-information devices. We also draw parallels to the QR geometry with the electrostatic potential being a linear combination of the terms proportional and inversely proportional to the square of the distance from the center \cite{Olendski2,Olendski3}. Hence, the present research, being a prerequisite for comparative analysis of the Dirichlet and Neumann BCs influence on the quantum information measures, simultaneously does the same for the two distinct shapes of the confining potential: one, where the motion is restricted by the hard walls to the finite space, whereas the second configuration permits the particle to be found at any point of the punctured at the origin boundless domain. Accordingly, it is believed that the results presented below will facilitate the growth of the annular structures with predefined quantum-information properties. QRs present a unique playground for the quantum-mechanical paradigm with paramount promises in both fundamental science and engineering \cite{Fomin1}; in particular, microring resonators are used for filtering, sensing and other nonlinear photonic applications \cite{Krasnokutska1}. In advanced quantum-information technologies, they are an essential ingredient of the platform for generation of the ultra-bright entangled-photon pairs \cite{Steiner1}. In addition, semiconductor QRs can be used, e.g., as single-gate qubits with switching times in the terahertz regime where a complete control of single-electron states in them is established by laser pulses of two-component polarization \cite{Rasanen1}. Accordingly, a comprehensive knowledge of both position and momentum functionals, which for the different substances might strongly depend on the BCs and the shape of the confining potential, is essential for such most advantageous manipulation.

Structure of the paper is as follows. In Sec.~\ref{sec_Model}, considered model is introduced and expressions for the position and momentum waveforms together with the energy spectrum are derived and analyzed; in particular, limiting cases of the thick and thin rings are discussed. In addition, a long-standing problem of the units of calculating measures of the continuous probability distributions is tackled too. Sec.~\ref{sec_measures} uses the results of the previous part to calculate Shannon and R\'{e}nyi entropies, Fisher information and Onicescu energy in terms of the geometry of the ring and AB flux. Special attention is devoted to the comparison between different models and their similarities and differences are pointed out. Sec.~\ref{sec_Conclusions} summarizes the findings and proposes future lines of research.

\section{Model and formulation}\label{sec_Model}
The structure we consider is a flat 2D annular region of the width $a$ and inner radius $r_0$ inside which the particle with mass $M$ and charge $q$ (for definiteness, we will talk about the electron, $M=m_e^*$, $q=-|e|$, with $m_e^*$ being an effective mass of the carrier in the crystal) is moving. The confining circles have the same center and the position wave function vanishes at them, what constitutes the Dirichlet BC. In addition, the AB flux $\phi_{AB}$ threads the QR plane at the origin. Regime $r_0\ll a$ describes the thick ring with the extreme $r_0=0$ degenerating it to the QD whereas the opposite limit $r_0\gg a$ is relevant for the thin QRs. Different variations of such geometry were theoretically considered before \cite{Peshkin1,Skarzhinskiy1,Afanasev1,Avishai1,Olendski4} with primary aim of investigating the flux influence on the energy spectrum, persistent currents \cite{Buttiker1} and magnetizations. Instead, our motivation lies in the analysis of the quantum-information functionals described in the Introduction and their dependence on the ring geometry and the AB field. But prior to starting the discussion, it is important to address the issue of the units in which the measures are calculated; namely, expressions for the Shannon entropies of the continuous $\mathtt{d}$-dimensional variables $\bf r$ and $\bf k$ show that they, contrary to their dimensionless discrete counterpart from Eq.~\eqref{ShannonDiscrete1}, are evaluated in units of the logarithm of the distance:
\begin{subequations}\label{ShannonDimensionless1}
\begin{align}\label{ShannonDimensionless1_rho}
S_{\rho_\mathtt{n}}&=\mathtt{d}\ln L+\overline{S}_{\rho_\mathtt{n}},\\
\label{ShannonDimensionless1_gamma}
S_{\gamma_\mathtt{n}}&=-\mathtt{d}\ln L+\overline{S}_{\gamma_\mathtt{n}},
\end{align}
\end{subequations}
with $L$ being some characteristic length of the system and the overline denoting a dimensionless $L$-independent term. As it follows from Eqs.~\eqref{ShannonDimensionless1}, the sum of the position and momentum functionals $S_t$ is a unitless scale-independent quantity what also is immediately seen from the uncertainty relation, Eq.~\eqref{ShannonInequality1}. If the description of the system contains more than one distance (e.g., the QR is characterized by its inner $r_0$ and outer $r_1=r_0+a$ radii), then mathematically either of them can be used as a logarithm argument with the dimensionless parts from Eqs.~\eqref{ShannonDimensionless1} being dependent on the ratios of the remaining lengths to that entering the logarithm. Since for our geometry the motion is confined to the region of the width $a$, physically it is the most natural to choose just it as a basic one. Accordingly, below, together with the regular quantities we, in order to simplify the exposition, often will simultaneously use their dimensionless counterparts; namely, all overlined spatial parameters and variables denote the ratios of the corresponding dimensional lengths to $a$; the energies, if necessary, will be represented in units of the ground-state energy of the 1D Dirichlet well of the same width, $E=\frac{\pi^2\hbar^2}{2m_e^*a^2}\overline{E}$; wave vectors -- in units of inverse ring width, $k=\overline{k}/a$; areas -- in units of $a^2$, etc. Different methods have been used for resolving the physical concerns about the long-standing equivocation of the units of entropies; for example, following Shannon \cite{Shannon2}, one can exponentiate the measures from Eqs.~\eqref{Shannon1} \cite{Srinivas1,Hall1} and work with the corresponding information-theoretic position and momentum lengths. Alternatively, by multiplying the logarithms' arguments in these definitions by the appropriate scale factor, one makes the functionals dimensionless \cite{Dodonov1,Matta1,Flores1,Nascimento1,Nascimento2} but, obviously, the values of $S_\rho$ and $S_\gamma$ depend in this case on the scale multiplier. Eqs.~\eqref{ShannonDimensionless1} suggest another way of handling the entropies; namely, since such singling out is always possible, one should consider dimensionless quantities $S_\rho-\mathtt{d}\ln L$ and $S_\gamma+\mathtt{d}\ln L$. Following this convention, below we stay focused on the analysis of
\begin{subequations}\label{ShannonDimensionless2}
\begin{align}\label{ShannonDimensionless2_rho}
\overline{S}_\rho&=S_\rho-2\ln a\\
\label{ShannonDimensionless2_gamma}
\overline{S}_\gamma&=S_\gamma+2\ln a,
\end{align}
\end{subequations}
which depend on the normalized inner radius $\overline{r}_0$ only. Of course, the same strategy applies to the R\'{e}nyi entropies too. With respect to the Fisher informations and disequilibria, one can identify their unitless parts according to
\begin{subequations}\label{FisherDimensionless1}
\begin{align}\label{FisherDimensionless1_rho}
I_\rho&=L^{-2}\overline{I}_\rho,\\
\label{FisherDimensionless1_gamma}
I_\gamma&=L^2\overline{I}_\gamma,
\end{align}
\end{subequations}
which is true for any number $\mathtt{d}$, and
\begin{subequations}\label{OnicescuDimensionless1}
\begin{align}\label{OnicescuDimensionless1_rho}
O_\rho&=L^{-\mathtt{d}}\,\overline{O}_\rho,\\
\label{OnicescuDimensionless1_gamma}
O_\gamma&=L^\mathtt{d}\,\overline{O}_\gamma,
\end{align}
\end{subequations}
what makes the corresponding products $I_\rho I_\gamma$ and $O_\rho O_\gamma$ dimensionless quantities. Also, in the wake of the above discussion, it is clear that the complexity $CGL$ in either space has no units either.

Axial symmetry of the system dictates the choice of the polar coordinates as the most convenient basis, ${\bf r}=(r,\varphi_{\bf r})$ and ${\bf k}=(k,\varphi_{\bf k})$. Then, the domain of the 2D Hamiltonian $\widehat{H}$ from Eq.~\eqref{Hamiltonian1} consists of doubly differentiable functions $\Psi$ that are defined inside the annulus and vanish at $r=r_0$ and $r=r_1$:
\begin{equation}\label{Domain1}
\mathcal{D}_\rho^{(2)}\left(\widehat{H}\right)=\left\{\Psi,{\bm\nabla}_{\bf r}^2\Psi\in\mathcal{L}^2(r_0\leq r\leq r_1,0\leq\varphi_{\bf r}<2\pi),\Psi(r_0,\varphi_{\bf r})=\Psi(r_1,\varphi_{\bf r})=0\right\}.
\end{equation}
In the symmetric gauge, the vector potential of the AB whisker has a nonzero azimuthal component only,
\begin{equation}\label{VectorPotential1}
{\bf A}=\left(0,\frac{\phi_{AB}}{2\pi r}\right);
\end{equation}
accordingly, the magnetic field is perpendicular to the QR plane, ${\bf B}({\bf r})=B(r){\bf z}$ (with ${\bf z}$ being a unit vector along the $z$-axis),  and is localized at  the origin only:
\begin{equation}\label{MagneticField1}
 B(r)=\phi_{AB}\delta(x)\delta(y)=\phi_{AB}\frac{\delta(r)}{\pi r},
\end{equation}
where $\delta(\xi)$ is a 1D $\delta$-function and $(x,y)$ are Cartesian coordinates with the same origin.

Schr\"{o}dinger equation for the function $\Psi\left(\overline{\phi}_{AB},\overline{r}_0;{\bf r}\right)$ reads:
\begin{equation}\label{WaveEquation1}
\frac{1}{r}\frac{\partial}{\partial r}\left(r\frac{\partial\Psi}{\partial r}\right)+\frac{1}{r^2}\frac{\partial^2\Psi}{\partial\varphi_{\bf r}^2}+2i\frac{\overline{\phi}_{AB}}{r^2}\frac{\partial\Psi}{\partial\varphi_{\bf r}}-\left(\frac{\overline{\phi}_{AB}}{r}\right)^2\Psi+\frac{2m_e^*E}{\hbar^2}\Psi=0.
\end{equation}
Here, dimensionless AB flux is expressed in units of the elementary flux quantum $\phi_0=h/|e|$:
\begin{equation}\label{ABdimensionless1}
\overline{\phi}_{AB}=\frac{\phi_{AB}}{\phi_0}.
\end{equation}
In Eq.~\eqref{WaveEquation1}, the variables are separated:
\begin{equation}\label{Separation1}
\Psi_{nm}(r,\varphi_{\bf r})=\frac{1}{\sqrt{2\pi}}e^{im\varphi_{\bf r}}{\cal R}_{nm}(r),
\end{equation}
where the combined quantum number $\mathtt{n}$ has been split into the principal $n$ and azimuthal $m$ indices, $n=1,2\ldots$, $m=0,\pm1,\ldots$. Radial dependence ${\cal R}_{nm}(r)\equiv{\cal R}_{nm}\left(\overline{\phi}_{AB},\overline{r}_0;r\right)$ obeys the equation
\begin{equation}\label{RadialEquation1}
\frac{d^2{\cal R}_{nm}}{dr^2}+\frac{1}{r}\frac{d{\cal R}_{nm}}{dr}+\left(\frac{2m_e^*E}{\hbar^2}-\frac{m_\phi^2}{r^2}\right){\cal R}_{nm}=0,
\end{equation}
$m_\phi=m+\overline{\phi}_{AB}$. To guarantee the orthonormalization, Eq.~\eqref{OrthoNormality1}, it also should satisfy:
\begin{equation}\label{OrthoNormality2}
\int_{r_0}^{r_1}r{\cal R}_{nm}(r){\cal R}_{n'm}(r)dr=\delta_{nn'}.
\end{equation}
Moreover, due to the Dirichlet BC, it vanishes at the edges,
\begin{equation}\label{RadialBC1}
{\cal R}(r_0)={\cal R}(r_1)=0.
\end{equation}
Then, it reads:
\begin{equation}\label{PositionRadial1}
{\cal R}_{nm}(r)=\frac{N_{nm}}{a}\left[Y_{|m_\phi|}\left(\pi\overline{E}_{nm}^{1/2}\overline{r}_0\right)J_{|m_\phi|}\left(\pi\overline{E}_{nm}^{1/2}\overline{r}\right)-J_{|m_\phi|}\left(\pi\overline{E}_{nm}^{1/2}\overline{r}_0\right)Y_{|m_\phi|}\left(\pi\overline{E}_{nm}^{1/2}\overline{r}\right)\right]
\end{equation}
with the normalization constant $N_{nm}$ being
\begin{equation}\label{Normalization1}
N_{nm}^{-2}=\frac{\overline{r}_1^2}{2}\left[Y_{|m_\phi|}\!\!\left(\pi\overline{E}_{nm}^{1/2}\overline{r}_0\!\right)\!J_{|m_\phi|+1}\!\left(\pi\overline{E}_{nm}^{1/2}\overline{r}_1\!\right)-J_{|m_\phi|}\!\!\left(\pi\overline{E}_{nm}^{1/2}\overline{r}_0\!\right)\!Y_{|m_\phi|+1}\!\left(\pi\overline{E}_{nm}^{1/2}\overline{r}_1\!\right)\right]^2-\frac{2}{\pi^4\overline{E}_{nm}}.
\end{equation}
Here, $J_\nu(x)$ and $Y_\nu(x)$ are Bessel functions of the first and second kind, respectively \cite{Abramowitz1}. Transcendental equation for finding the energies $\overline{E}_{nm}$ reads \cite{Skarzhinskiy1,Afanasev1}:
\begin{equation}\label{EigenValueEquation1}
J_{|m_\phi|}\left(\pi\overline{E}_{nm}^{1/2}\overline{r}_0\right)Y_{|m_\phi|}\left(\pi\overline{E}_{nm}^{1/2}\overline{r}_1\right)-Y_{|m_\phi|}\left(\pi\overline{E}_{nm}^{1/2}\overline{r}_0\right)J_{|m_\phi|}\left(\pi\overline{E}_{nm}^{1/2}\overline{r}_1\right)=0.
\end{equation}
From here, it is immediately seen that the energies obey the following relations:
\begin{equation}\label{ABrelation1}
E_{nm}\!\left(\overline{\phi}_{AB},\overline{r}_0\right)=E_{n,m-1}\!\left(\overline{\phi}_{AB}+1,\overline{r}_0\right)=E_{n,m+1}\!\left(\overline{\phi}_{AB}-1,\overline{r}_0\right),
\end{equation}
what allows to limit a consideration of the range of variation of the normalized AB field to
\begin{equation}\label{Abrange1}
-\frac{1}{2}\leq\overline{\phi}_{AB}\leq\frac{1}{2},
\end{equation}
and at the edges of this interval the following radius-independent relations hold:
\begin{subequations}\label{ABrelations2}
\begin{align}\label{ABrelations2Minus}
\overline{E}_{nm}\left(-\frac{1}{2},\overline{r}_0\right)&=\overline{E}_{n,-m+1}\left(-\frac{1}{2},\overline{r}_0\right)\\
\label{ABrelations2Plus}
\overline{E}_{nm}\left(\frac{1}{2},\overline{r}_0\right)&=\overline{E}_{n,-m-1}\left(\frac{1}{2},\overline{r}_0\right).
\end{align}
\end{subequations}
Eq.~\eqref{EigenValueEquation1} also means that the energy of the $m=0$ orbital is an even function of the flux:
\begin{equation}\label{ABsymmetry1}
E_{n,0}\!\left(-\overline{\phi}_{AB},\overline{r}_0\right)=E_{n,0}\!\left(\overline{\phi}_{AB},\overline{r}_0\right).
\end{equation}

Even though physicists and mathematicians have been dealing with the roots of the equations of the type from Eq.~\eqref{EigenValueEquation1} since, at least, the end of the 19th century \cite{McMahon1} with a lot of knowledge being accumulated over the years \cite{Abramowitz1,Lowan1,Kline1,Dwight1,Waldron1,Jahnke1,Laslett1,Gunston1,Cochran1,Cochran2,Willis1,Marcuvitz1,Cochran3}, nowadays its miscellaneous properties continue to attract the researchers \cite{Bobkov1,Guo1}. Generally, it can be handled numerically only but in the extreme regimes of the thick and thin rings its analytic asymptotic solutions are:
\begin{equation}\label{Asymptote1}
\overline{E}_{nm}=\left\{\begin{array}{cc}
\left(\frac{j_{|m_\phi|,n}}{\pi}\right)^2(1-2\overline{r}_0)-\left\{\begin{array}{cc}
\frac{j_{0n}}{\pi}\frac{Y_0(j_{0n})}{J_1(j_{0n})}\frac{1}{\ln(j_{0n}\overline{r}_0)},&m_\phi=0\\
\frac{2j_{|m_\phi|,n}}{\pi\Gamma(|m_\phi|)\Gamma(|m_\phi|+1)}\frac{Y_{|m_\phi|}(j_{|m_\phi|,n})}{J'_{|m_\phi|}(j_{|m_\phi|,n})}\left(\frac{j_{|m_\phi|,n}\overline{r}_0}{2}\right)^{2|m_\phi|},&m_\phi\neq0
\end{array}\right\}
,&\overline{r}_0\rightarrow0\\
n^2+\frac{4|m_\phi|^2-1}{4\pi^2}\frac{1}{\overline{r}_0^2},&\overline{r}_0\rightarrow\infty,
\end{array}\right.
\end{equation}
with $j_{\nu n}$ being $n$th root of the Bessel function $J_\nu(x)$ \cite{Abramowitz1} and $\Gamma(x)$ being $\Gamma$-function \cite{Abramowitz1}. This expression shows that for the vanishingly small inner radius the energies approach those of the QD \cite{Olendski1}, as expected. In the opposite regime of the very thin QR, they tend to those of the 1D Dirichlet well, which is not surprising either. Note that, for the latter asymptote, the influence of the magnetic flux fades inversely proportional to the square of the radius what is elementary explained by the increasing distance between the AB whisker and the region where the electron is moving.

\begin{figure}
\centering
\includegraphics[width=\columnwidth]{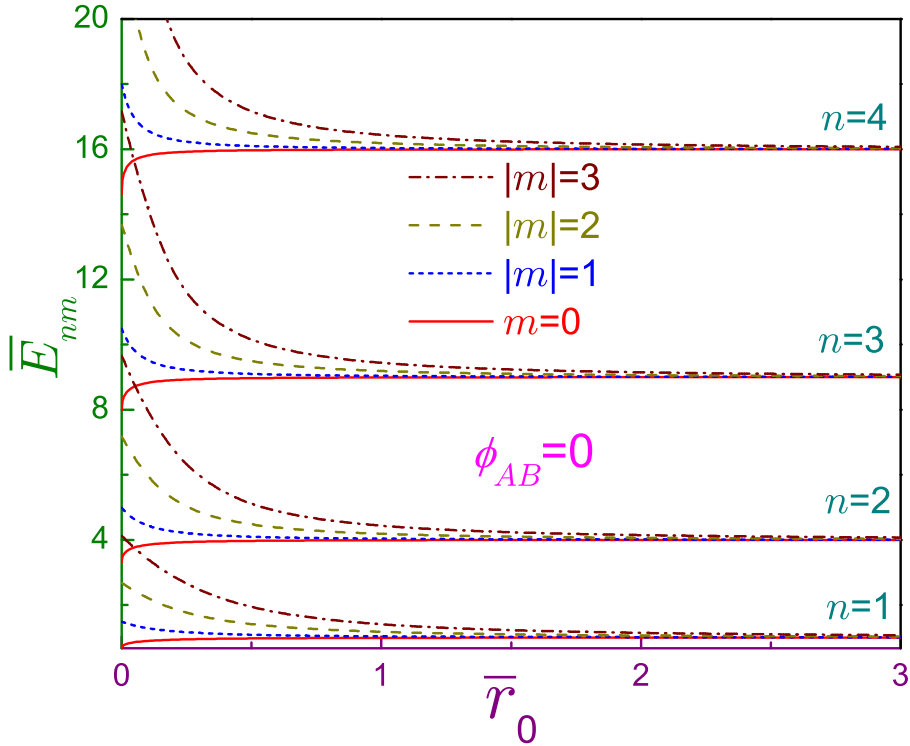}
\caption{\label{Fig_Energies1}
Dimensionless energies $\overline{E}_{nm}$ as functions of the unitless inner radius $\overline{r}_0$ at zero AB flux. Rotationally symmetric, $m=0$, orbitals are depicted by the solid lines, dashed curves are for the states with $|m|=1$, dotted ones -- for  $|m|=2$, and dash-dotted lines are for the levels with  $|m|=3$.}
\end{figure}

Fig.~\ref{Fig_Energies1} depicts spectrum evolution with inner radius variation. From their disc values at $\overline{r}_0=0$ \cite{Olendski1}, the energies monotonically approach at $\overline{r}_0\rightarrow\infty$ the $m$-independent 1D magnitudes of $n^2$. In this path to the degeneracy, the rotationally symmetric levels, $m=0$, lie always below the corresponding limit with all other orbitals located above it, as it follows from Eq.~\eqref{Asymptote1} and is vividly exemplified in the figure. For the greater $|m|$, the asymptote is reached at the longer inner radius.

\begin{figure}
\centering
\includegraphics[width=\columnwidth]{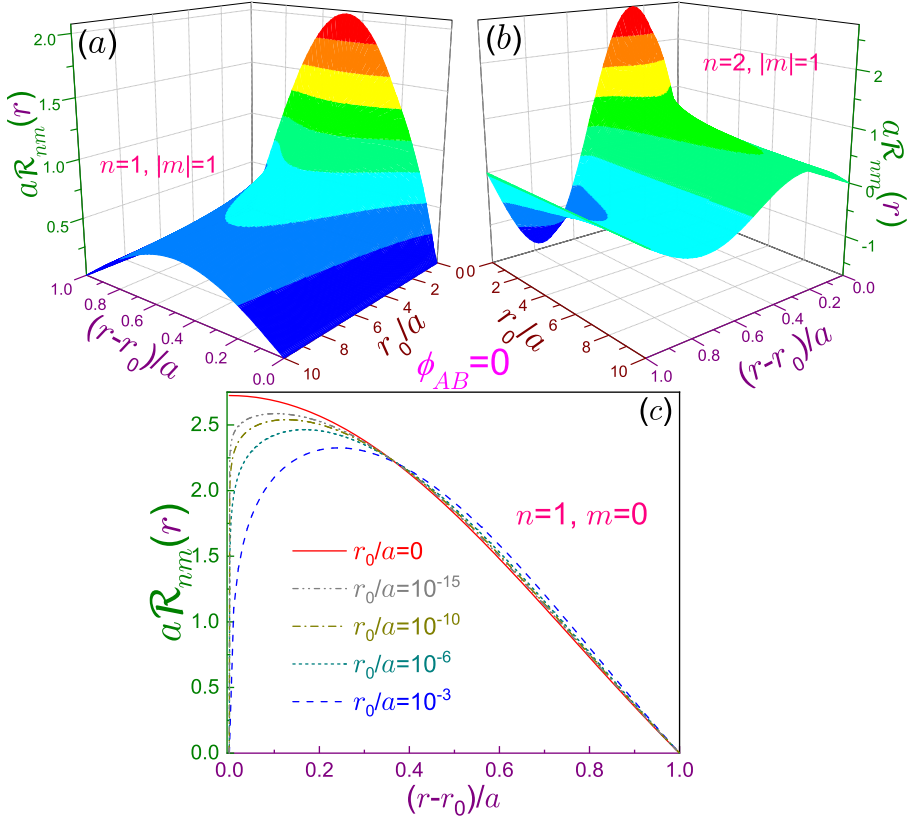}
\caption{\label{Fig_PositionWaveforms1}
AB-free dimensionless radial waveforms $a{\cal R}_{nm}(r)$ in terms of the scaled inner radius $r_0/a$ for (a) $n=1$, $|m|=1$ and (b) $n=2$, $|m|=1$. (c) Ground-state wave function $a{\cal R}_{10}(r)$ for several small and ultra small radii $\overline{r}_0$. For comparison, the QD dependence, $r_0=0$, is shown too.}
\end{figure}

In Fig.~\ref{Fig_PositionWaveforms1}, transformations of the position functions ${\cal R}_{nm}(r)$ are shown when the QR shape changes from the thick to the thin geometry. Increasing inner radius $\overline{r}_0$ flattens the slope of the waveforms. Utilizing properties of the Bessel functions \cite{Abramowitz1}, one corroborates this feature by the following asymptote:
\begin{equation}\label{Asymptote2}
{\cal R}_{nm}(r)=\frac{1}{r_0^{1/2}}\,\psi_n^{(1D)}(r-r_0),\quad r_0\rightarrow\infty,
\end{equation}
where $\psi_n^{(1D)}(x)$ is the function of the $n$th level of the 1D Dirichlet well of the width $a$:
\begin{equation}\label{Radial1D}
\psi_n^{(1D)}(x)=\sqrt{\frac{2}{a}}\sin\left(n\pi\frac{x}{a}\right),\quad0\leq x\leq a.
\end{equation}
Note that in this regime, the dependencies on the magnetic quantum number and the AB flux have been lost, as was the case for the energy spectrum too. Next, a transformation from the QD, $r_0=0$, to the thick QR with very small but nonzero inner radius does not have a conspicuous effect on the $m\neq0$ levels, as windows (a) and (b) exemplify, since for the former geometry their position function vanishes at the center of the disc. However, since the QD axially symmetric orbitals, $m=0$, are the only ones that have a finite probability of finding a particle at the origin \cite{Olendski1}, this change of the topology of the system from simply to doubly connected domain drastically influences the evolution of the corresponding waveforms. Panel (c) exhibits ground-state function $R_{10}(r)$ at several small and extremely small $r_0$. For comparison, a dot dependence is shown too. At the close vicinity of the inner edge, there are deviations from the disc behavior since the inner Dirichlet BC forces the waveform to fade to zero at $r=r_0$. A spatial range of the noticeable deflection from the QD dependence increases with the radius $r_0$. For the thin rings, their $m=0$ states evolve similar to all other orbitals.

Knowledge of the energy spectrum and position wave functions paves the way to finding all other characteristics of the structure; for example, calculation of a root-mean-square radius 
\begin{equation}\label{RootMeanSquare1}
\mathsf{r}=\sqrt{\int_{{\cal D}_\rho^{(\mathtt{d})}}r^2\rho({\bf r})d{\bf r}}
\end{equation}
reveals that the increase of $|m|$ brings from below its normalized value closer and closer to $\overline{r}_1$ what means, similar to the QD \cite{Olendski1}, a stronger localization of the electron at the outer surface. Contrary, for the Volcano-type potential \cite{Olendski2}
\begin{equation}\label{Volcano1}
V(\mathbf{r})\equiv V_V(r)=\frac{1}{2}M\omega_0^2r^2+\frac{\hbar^2}{2Mr^2}\,\mathtt{a}-\hbar\omega_0\mathtt{a}^{1/2},
\end{equation}
where frequency $\omega_0$ defines a steepness of the confining in-plane outer interface of the QR and positive dimensionless constant $\mathtt{a}$ describes a strength of the repulsive potential near the origin, this radius is an unrestrictedly increasing function of the absolute value of the azimuthal quantum number \cite{Olendski9}:
\begin{equation}\label{Volcano2}
\mathsf{r}_V=\left[\left(2n+\sqrt{m^2+\mathtt{a}}-1\right)\frac{\hbar}{M\omega_0}\right]^{1/2}.
\end{equation}
Obviously, such unalike behavior is due to the different character of motion: for the Dirichlet geometry, it is strictly limited by the hard walls to the finite domain $r_0\leq r\leq r_0+a$ whereas for the Volcano disc the particle can be found at the arbitrary small as well as unboundedly large distances from the QR center. As we shall see later, this distinction causes dissimilar dependencies of, e.g., Shannon entropy on $m$ and the AB flux. Next, an expression for the current density ${\bf j}({\bf r})$ \cite{Landau1} yields for its azimuthal component:
\begin{equation}\label{CurrentDensity1}
j_{\varphi_{nm}}=-\frac{1}{2\pi}\frac{|e|\hbar}{m_e^*}\frac{m_\phi}{r}{\cal R}_{nm}^2(r),
\end{equation}
what means that the total persistent current of the particular orbital flowing along the ring is calculated as:
\begin{equation}\label{Current1}
J_{nm}=-\frac{1}{2\pi}\frac{|e|\hbar}{m_e^*}\,m_\phi\int_{r_0}^{r_1}\frac{dr}{r}{\cal R}_{nm}^2(r).
\end{equation}
Hellmann-Feynman theorem adds two more equivalent expressions:
\begin{equation}\label{Current2}
J_{nm}=-\frac{|e|}{h}\frac{\partial E_{nm}}{\partial m}=-\frac{\partial E_{nm}}{\partial\phi_{AB}}.
\end{equation}
Similarly, the one-particle magnetization operator \cite{Grochol1}
\begin{equation}\label{Magnetization1}
\widehat{{\bf M}}=\frac{1}{2}{\bf r}\times\widehat{\bf j}({\bf r})
\end{equation}
leads to the following $n$-independent magnetization of the quantum state:
\begin{equation}\label{Magnetization2}
{\bf M}=-\frac{|e|\hbar}{m_e^*}\,m_\phi{\bf z}.
\end{equation}
However, the most crucial step in continuation of our research is a calculation of the momentum functions, which, according to Eq.~\eqref{Fourier1_1}, are:
\begin{equation}\label{MomentumFunction1}
\Phi_{nm}({\bf k})\equiv\Phi_{nm}(k,\varphi_{\bf k})=\frac{1}{(2\pi)^{3/2}}\int_0^{2\pi}d\varphi_{\bf r}\int_{r_0}^{r_1}drr{\cal R}_{nm}(r)e^{i[m\varphi_{\bf r}-kr\cos(\varphi_{\bf r}-\varphi_{\bf k})]}.
\end{equation}
Angular integration is performed identically to the QD geometry \cite{Olendski1} what means that in the momentum space the variables are separated too:
\begin{equation}\label{Separation2}
\Phi_{nm}(k,\varphi_{\bf k})=\frac{(-i)^{|m|}}{\sqrt{2\pi}}e^{im\varphi_{\bf k}}{\cal K}_{nm}(k),
\end{equation}
and radial dependence ${\cal K}_{nm}(k)\equiv{\cal K}_{nm}\left(\overline{\phi}_{AB},\overline{r}_0;k\right)$ reads:
\begin{equation}\label{MomentumRadial1}
{\cal K}_{nm}\left(\overline{\phi}_{AB},\overline{r}_0;k\right)=\int_{r_0}^{r_1}J_{|m|}(kr){\cal R}_{nm}\left(\overline{\phi}_{AB},\overline{r}_0;r\right)rdr.
\end{equation}
For the AB-free case, the integral in Eq.~\eqref{MomentumRadial1} is calculated analytically \cite{Prudnikov1}:
\begin{align}
&{\cal K}_{nm}(0,\overline{r}_0;k)=a\frac{N_{nm}}{\overline{k}^2-\pi^2\overline{E}_{nm}}\left\{\frac{2}{\pi}J_{|m|}(r_0k)\right.\nonumber\\
\label{MomentumRadial2}
&\left.-\pi\overline{E}^{1/2}\overline{r}_1\left[Y_{|m|}\!\!\left(\pi\overline{E}_{nm}^{1/2}\overline{r}_0\!\right)\!J_{|m|+1}\!\left(\pi\overline{E}_{nm}^{1/2}\overline{r}_1\!\right)-J_{|m|}\!\!\left(\pi\overline{E}_{nm}^{1/2}\overline{r}_0\!\right)\!Y_{|m|+1}\!\left(\pi\overline{E}_{nm}^{1/2}\overline{r}_1\!\right)\right]J_{|m|}(r_1k)\right\}.
\end{align}
However, at the nonvanishing $\phi_{AB}$, due to the absence in the known literature of the explicit expressions of the integral from Eq.~\eqref{MomentumRadial1}, its direct numerical quadrature has been performed. Eq.~\eqref{MomentumRadial2} shows that, similar to the QD \cite{Olendski1}, the azimuthally symmetric levels, $m=0$, are the only ones that at the finite $r_0$ allow the particle to have zero momentum.

\begin{figure}
\centering
\includegraphics[width=\columnwidth]{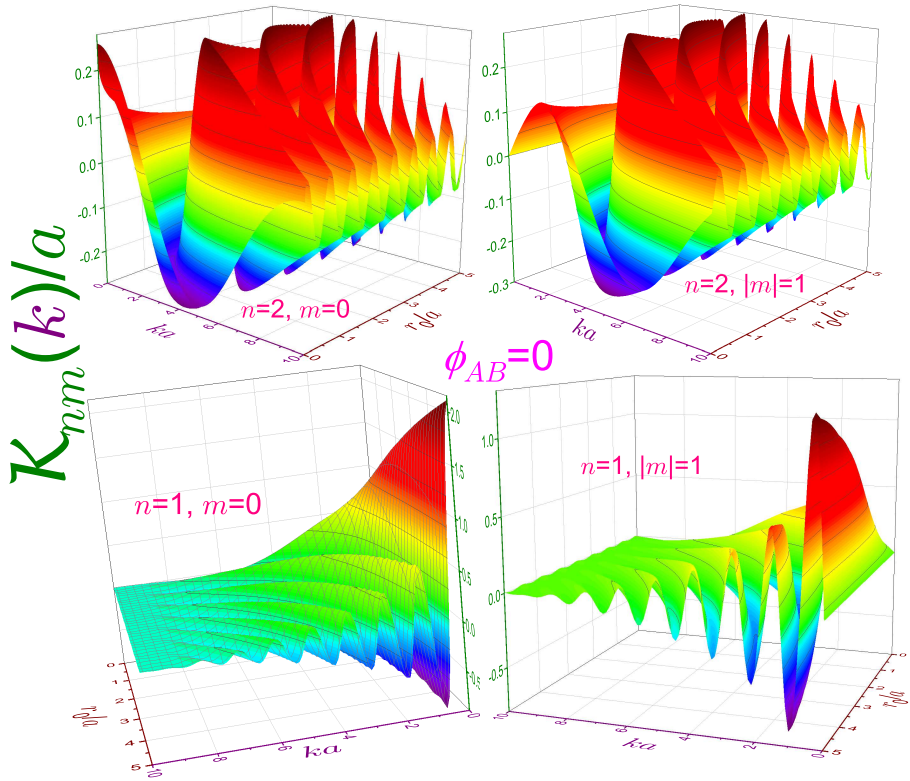}
\caption{\label{Fig_MomentumWaveforms1}
AB-free dimensionless radial momentum functions ${\cal K}_{nm}(k)/a$ in terms of the scaled inner radius $\overline{r}_0$ for several quantum numbers $n$ and $m$ denoted in each of the corresponding panel. Signs are chosen such that, e.g., ${\cal K}_{n0}(0)>0$. For better viewing, each subplot has its own perspective and distinct vertical range.}
\end{figure}

Fig.~\ref{Fig_MomentumWaveforms1} depicts AB-free momentum functions for the two smallest $n$ and $|m|$ with the geometry evolving from the QD to the thin annulus.  Similar patterns were observed for the other levels too. Dependence ${\cal K}_{nm}(k)$ at any QR shape exhibits an infinite number of maxima and minima. At $r_0=0$, among all the orbitals with the same magnetic quantum number, the lowest energy state, $n=1$, possesses the highest extremum which for the axially symmetric orbitals is achieved at the zero wave vector and for the growing $|m|$ it decreases shifting at the same time to the greater $k$ \cite{Olendski1}. The increase of $\overline{r}_0$ pushes the extrema to the smaller momenta simultaneously shrinking the interval between them. For the large $\overline{r}_0$, the radial function from Eq.~\eqref{MomentumRadial2} degenerates to
\begin{subequations}\label{MomentumRadialAsymptote1}
\begin{align}\label{MomentumRadialAsymptote1_a}
{\cal K}_{nm}(0,\overline{r}_0;k)&=a\frac{2^{1/2}\pi n}{\overline{k}^2-\pi^2n^2}\,\overline{r}_0^{1/2}\left[J_{|m|}(kr_0)-(-1)^nJ_{|m|}(kr_1)\right],\quad\overline{r}_0\gg1.
\intertext{From here, it follows that for the $m=0$ orbitals with odd (even) principal number the probability of the $k=0$ state grows linearly (tends to zero) with the elongating inner radius:}
\label{MomentumRadialAsymptote1_b}
{\cal K}_{n0}(0,\overline{r}_0;0)&=\left\{\begin{array}{cc}
-a\frac{2^{3/2}}{\pi n}\,\overline{r}_0^{1/2},&n\,{\rm odd}\\
0,&n\,{\rm even}
\end{array}\right\},\quad\overline{r}_0\gg1.
\intertext{This increase of $|{\cal K}_{10}(0;0)|$ and, in agreement with Eq.~\eqref{MomentumRadialAsymptote1_b}, a decrease of the zero-momentum maximum for the $n=2$ levels are vividly seen in Fig.~\ref{Fig_MomentumWaveforms1}. In addition, the limit $kr_0\gg1$ of Eq.~\eqref{MomentumRadialAsymptote1_a} helps to qualitatively explain  mentioned above growth of the frequency of the fading oscillations:}
\label{MomentumRadialAsymptote1_c}
{\cal K}_{nm}(0,\overline{r}_0;k)&=a\frac{4\pi^{1/2}n}{\overline{k}^{5/2}}\left\{\begin{array}{cc}
\cos\frac{\overline{k}}{2}\cos kr_0,&n\,{\rm odd}\\
\sin\frac{\overline{k}}{2}\sin kr_0,&n\,{\rm even}
\end{array}
\right\},\quad kr_0\gg1,\,\overline{r}_0\gg1.
\end{align}
\end{subequations}
In this first approximation, the magnitudes of the extrema are not influenced by the inner radius but the exact results from Fig.~\ref{Fig_MomentumWaveforms1} show that for the ground states  (and essentially at any odd principal index) of each constant $|m|$ they climb higher whereas for $n=2$ the amplitude of each maximum or minimum initially gets larger with $r_0$ and, after reaching its peak, it slowly diminishes. For the QD extrema that were located closer to $k=0$, this process develops at the smaller radius. Thus, the momentum functions of the levels with the odd principal numbers are characterized by the growing dominance (especially for $n=1$) of the maximum achieved at the smallest $k$ (which is zero for $|m|=0$) and even $n$ states exhibit a series of extrema of comparable amplitudes which at the elongating $\overline{r}_0$ are shifted to the smaller wave vectors.

\section{Quantum-information measures}\label{sec_measures}
Armed with understanding of the wave functions $\Psi({\bf r})$ and $\Phi({\bf k})$ behavior, one is able to compute the corresponding densities, Eq.~\eqref{Densities1}, and accordingly, to investigate quantum-information measures. As a first part of this analysis, the results are presented for the field-free geometry and next, AB flux influence is discussed too.
\subsection{AB-free geometry}\label{Subsec_ABfree}
\subsubsection{Shannon entropy}\label{Subsec_ShannonABfree}
\begin{figure}
\centering
\includegraphics[width=\columnwidth]{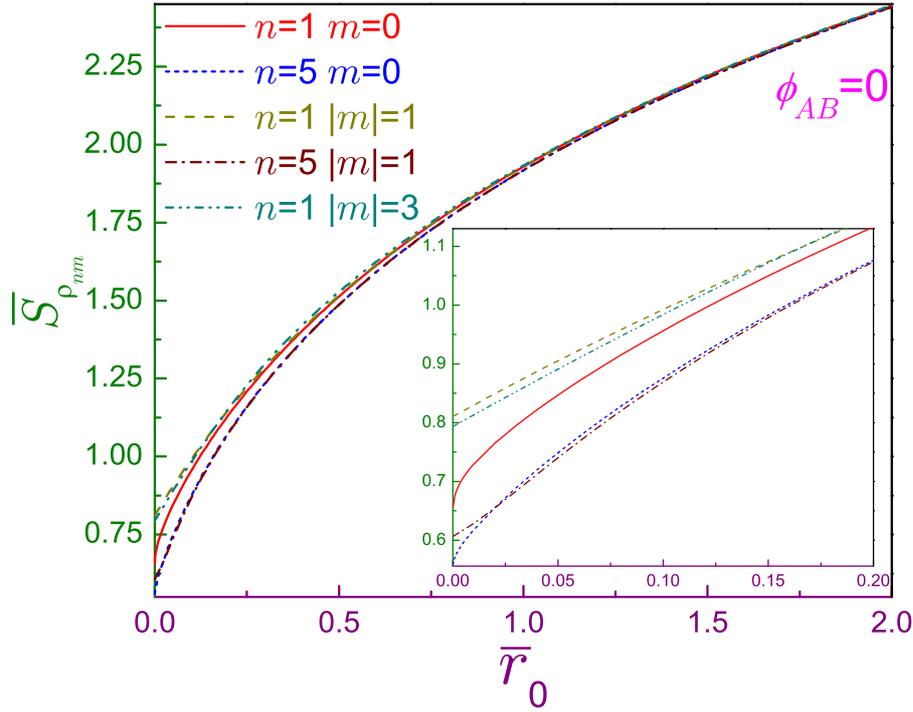}
\caption{\label{Fig_ShannonPosition1}
Position Shannon entropies $\overline{S}_{\rho_{nm}}$ for several $n$ and $m$ versus inner radius $\overline{r}_0$. Inset shows an enlarged view at the small $\overline{r}_0$.}
\end{figure}
Fig.~\ref{Fig_ShannonPosition1} shows evolution of the position Shannon entropy with the inner radius. From their QD values \cite{Olendski1}, the dimensionless functionals grow with $\overline{r}_0$ and at the large values of the latter logarithmically depend on it:
\begin{equation}\label{ShannonPositionAsymptote1}
\overline{S}_{\rho_{nm}}\rightarrow\ln(4\pi\overline{r}_0)-1,\quad\overline{r}_0\rightarrow\infty.
\end{equation}
This relation, which can be easily derived from the asymptotic shape of the waveform, Eq.~\eqref{Asymptote2}, and Fig.~\ref{Fig_ShannonPosition1} manifest that the position measure in this regime is the same for all orbitals: right-hand side of Eq.~\eqref{ShannonPositionAsymptote1} is independent of both indices $n$ and $m$. Figure shows that already at $\overline{r}_0\gtrsim1$ the difference between position Shannon quantities is quite small and decreases as the ring gets  thinner. Physically, entropy growth is construed as a shrinking of our knowledge about the particle location what is explained by the increasing area $\overline{A}$ of the annulus:
\begin{equation}\label{Area1}
\overline{A}=\pi(2\overline{r}_0+1).
\end{equation}
One has to note that, similar to the QD \cite{Olendski1}, the QR position measure is a nonmonotonic function of the azimuthal quantum number: this is especially vividly seen from the inset of Fig.~\ref{Fig_ShannonPosition1}, which demonstrates that it might persist at any shape of the ring, as for $n=1$, or emerges with variation of the inner radius, see dotted and dash-dotted lines for $n=5$. At the fixed $\overline{r}_0$, the limit of the extremely large $|m|$ squeezes the particle to the outer wall what means that its position is almost  perfectly known leading to the extremely huge negative $\overline{S}_\rho$. Contrary, the Volcano ring is characterized by the monotonically growing with $|m|$ position Shannon entropies that in the same limit behave as $\ln|m|^{1/2}$ \cite{Olendski2}. Such opposite behavior is a consequence of the different shape of the confining potential, as already clarified in Sec.~\ref{sec_Model}.

\begin{figure}
\centering
\includegraphics[width=\columnwidth]{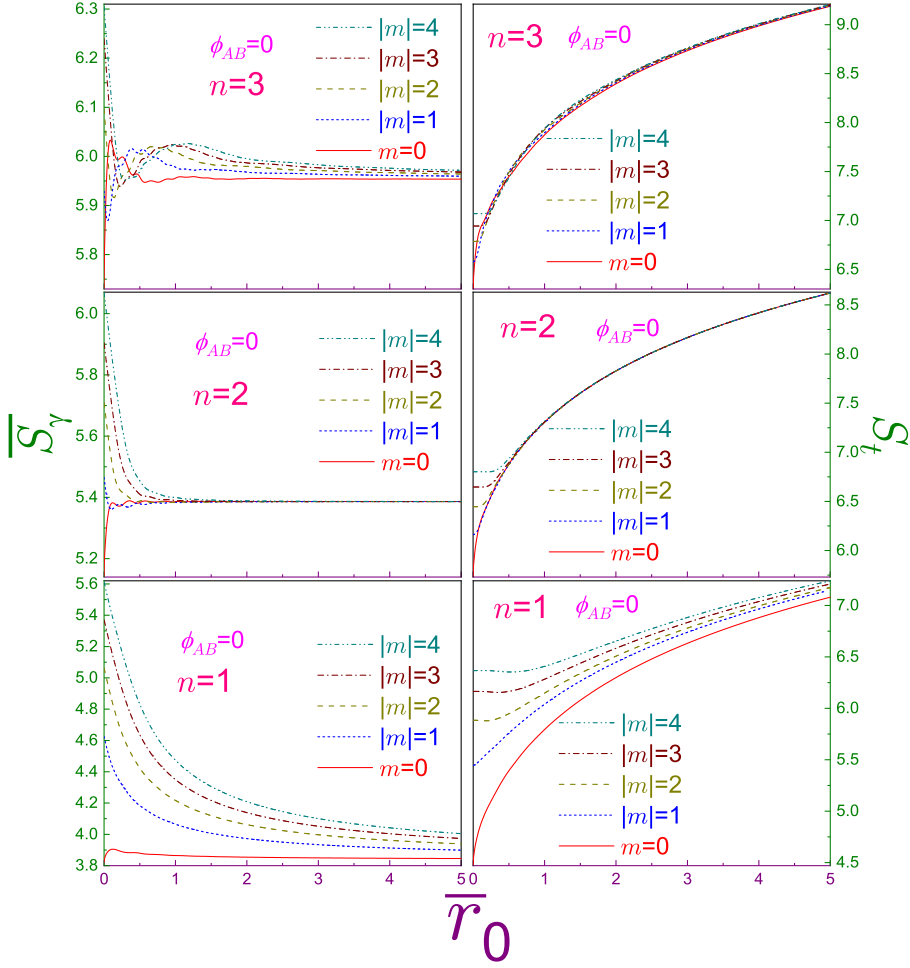}
\caption{\label{Fig_ShannonMomentumTotal1}
Dimensionless momentum Shannon functionals $\overline{S}_{\gamma_{nm}}$ (left windows) and total Shannon entropies $S_t$ (right panels) versus radius $\overline{r}_0$, where lower figures show the states with $n=1$ , middle subplots -- with $n=2$ and upper panels -- $n=3$. Solid curves are for $m=0$, dotted lines -- for $|m|=1$, dashed dependencies -- for $|m|=2$, dash-dotted ones -- for $|m|=3$, and dash-dotted ones -- for $|m|=4$. Note different ranges of each of the subplots.}
\end{figure}

As the inner radius grows from zero, the momentum Shannon entropies, in general, evolve non-monotonically with it; for example, the ground orbital exhibits a maximum of $3.905$ achieved at $\overline{r}_0\simeq0.124$ followed by the series of less noticeable extrema whereas the functionals of the other levels of the same subband, $n=1$, continuously decrease. As the left panels of Fig.~\ref{Fig_ShannonMomentumTotal1} exhibit, the irregular structures of $\overline{S}_{\gamma_{nm}}-\overline{r}_0$ characteristics become more conspicuous for the higher $n$ when the entropies of the states with the larger $|m|$ become more chaotic at the small and moderate ratio $r_0/a$. For the thin rings, $\overline{r}_0\gg1$, the amount of information about the particle wave vector ceases to depend on the radius: functionals of the different magnetic quantum numbers come closer and closer to the flat asymptotes whose magnitudes increase with the principal index. For the even $n$, these limiting values are approached at the smaller $\overline{r}_0$, what is attributed to the shape of the corresponding waveform, Eq.~\eqref{MomentumRadialAsymptote1_a}.

Sums of the two measures are depicted in the right panels of Fig.~\ref{Fig_ShannonMomentumTotal1}. Behavior of the total entropy is a consequence of the interaction of its constituents. Irregularities of the momentum functionals at the small and moderate radii are overshadowed by the robust growth of their position counterparts: except the states with $|m|=3$ and $|m|=4$ in the lowest two windows, the narrow regions at $\overline{r}_0\lesssim1$ of the tiny negative derivatives $\partial S_t/\partial{\overline{r}_0}$, due to their littleness, are not resolved in the scales of the figure. For the thin rings, the growth of the total entropy has the identical rate for each subband: the loss of the position information with increasing radius is the same for any orbital whereas $\overline{r}_0$-independent knowledge of the wave vector depends on the principal quantum number only. Needless to say that the Beckner-Bia{\l}ynicki-Birula-Mycielski \cite{Bialynicki1,Beckner1} uncertainty relation~\eqref{ShannonInequality1}, with its right-hand side for our geometry of $\mathtt{d}=2$ being $4.289459\ldots$, which held true by the QD \cite{Olendski1}, is all the more satisfied by the ring of the arbitrary nonzero radius.

\subsubsection{Fisher information}\label{Subsec_FisherABfree}
\begin{figure}
\centering
\includegraphics[width=0.7\columnwidth]{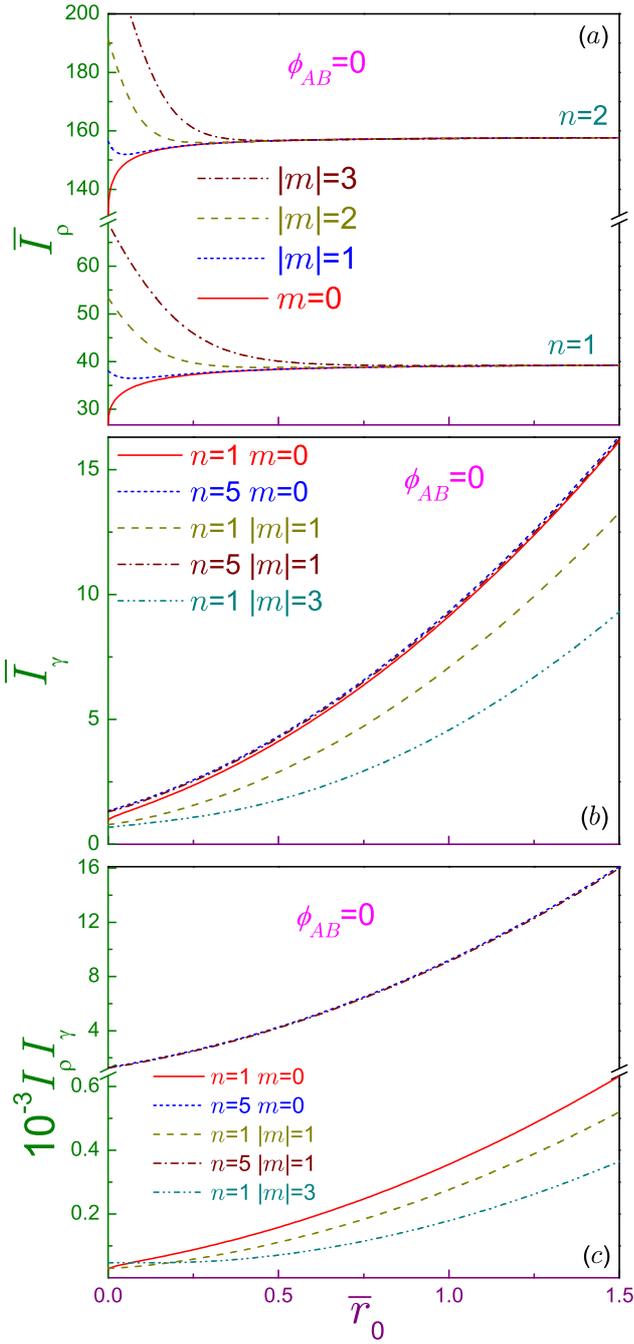}
\caption{\label{Fig_Fisher1}
Dimensionless (a) position $\overline{I}_\rho$ and (b) momentum $\overline{I}_\gamma$ Fisher informations together with their product $I_\rho I_\gamma$ [panel (c)] in terms of the normalized inner radius $\overline{r}_0$ for several quantum numbers $n$ and $m$ denoted in each subplot. Note vertical line breaks from $69$ to $131$ in the upper window and from $0.634$ to $1.21$ for the lower panel and different scales before and after the break. For the product of two informations, in subplot (c) a lowering factor $10^{-3}$ is used.}
\end{figure}

As an introduction to this part of the research, one needs to mention that, for the Volcano disc, $I_\rho$ does not depend on the magnetic quantum number and the AB flux \cite{Olendski2}. Position Fisher informations dependence on the Dirichlet ring geometry that is shown in panel (a) of Fig.~\ref{Fig_Fisher1} basically repeats the energy behavior, Fig.~\ref{Fig_Energies1}, what is explained by the proportionality relation between the position functional and kinetic energy of the system \cite{Sears1}; in particular, for the very thin structures they collapse into the subbands for each of which the  corresponding measures have the same principal index $n$. The first difference observed between $\overline{E}_{nm}$ and $\overline{I}_{\rho_{nm}}$ is the fact that the latters at $|m|\neq0$ are not monotonic functions of $\overline{r}_0$: before approaching at $\overline{r}_0\rightarrow\infty$ their 1D limits of $4\pi^2n^2$ \cite{Olendski5}, at some finite radius they reach a global minimum. One should also point out that the functionals touch their asymptotes at the smaller radii as compared to the energies.

Since Fisher informations are measures of the oscillating structure of the corresponding probability distribution functions and since, as discussed in the previous section, the waveforms ${\cal K}_{nm}(k)$ vary faster at the longer $\overline{r}_0$, momentum components $\overline{I}\gamma$ grow with the latter. This rise, which at the very thin rings approaches almost quadratic dependence, is depicted in Fig.~\ref{Fig_Fisher1}(b). As a result, the products of the two informations exhibit monotonically increasing dependence too, panel (c).

\subsubsection{Disequilibrium and complexity $e^SO$}\label{Subsec_OnicescuABfree}
\begin{figure}
\centering
\includegraphics[width=0.7\columnwidth]{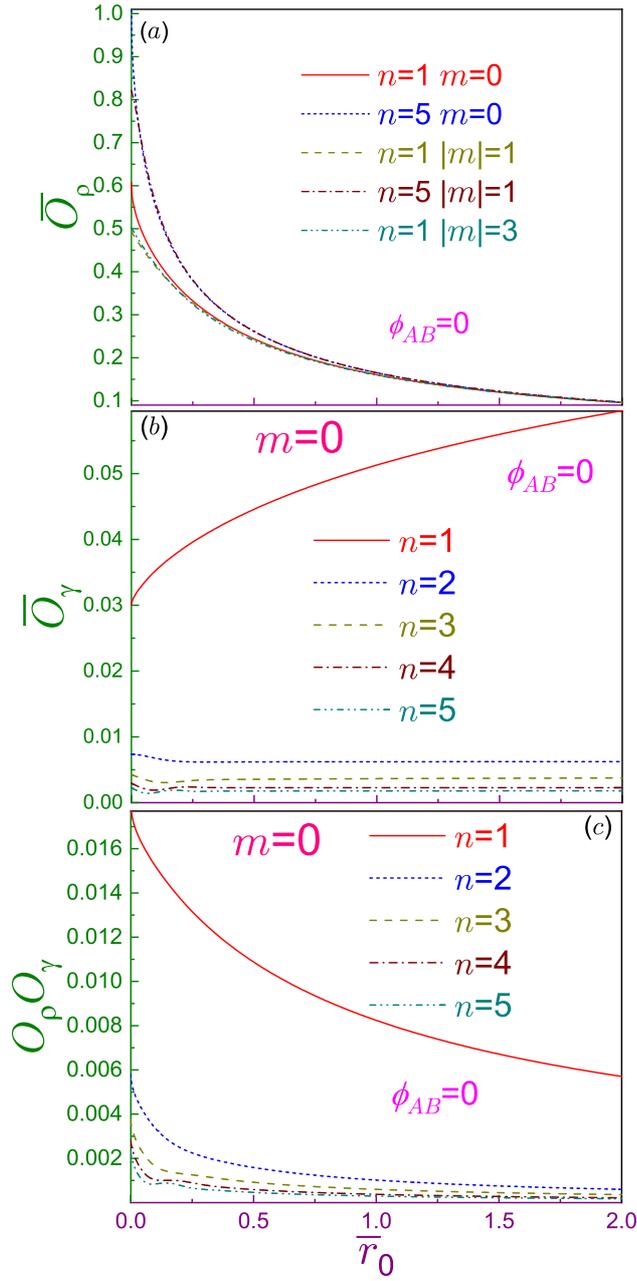}
\caption{\label{Fig_Onicescu1}
Dimensionless (a) position $\overline{O}_\rho$ and (b) momentum $\overline{O}_\gamma$ disequilibria together with their product $O_\rho O_\gamma$ [panel (c)] in terms of the normalized inner radius $\overline{r}_0$ for several quantum numbers $n$ and $m$ denoted in each subplot. Note different vertical ranges and scales for each subplot.}
\end{figure}

Position disequilibria monotonically decrease with the elongating radius, as is exemplified by Fig.~\ref{Fig_Onicescu1}(a), and at the large $\overline{r}_0$ they, independently of the orbital, follow the rule:
\begin{equation}\label{OnicescuPositionAsymptote1}
\overline{O}_{\rho_{nm}}\rightarrow\frac{3}{4\pi\overline{r}_0},\quad\overline{r}_0\rightarrow\infty.
\end{equation}
Physically, lessening of the Onicescu functional corresponds to the transformation to the more uniform distribution, which, according to Eq.~\eqref{Asymptote2}, really does take place for the position waveform at the thinner QR.

During the discussion of the momentum functions it was pointed out that the $n=1$ levels are characterized by the growing ascendancy of the extremum at the lowest wave vector. This increase of nonuniformity pushes the corresponding Onicescu functional higher, as shown by the solid line in panel (b). On the other hand, since other functions of the same $|m|$ family have a series of comparable peaks with the magnitudes smaller than the dominant one of the $n=1$ state, their momentum disequilibria do not vary considerably with $\overline{r}_0$ what implies the independence of the degree of the wave vector homogeneity from the radius. Let us also remark that the difference between the $n\neq1$ measures decreases with the principal quantum number with their $n=1$ counterpart being separated from them by the  widest gap.

The rate of growth of the ground-state momentum component is much slower than the decrease of corresponding position part; as a result, the ground-state product $O_\rho O_\gamma$ decreases as the QR becomes thinner. For the $n\neq1$ levels,  $O_\rho O_\gamma$ approaches zero much faster, as illustrated by Fig.~\ref{Fig_Onicescu1}(c).

\begin{figure}
\centering
\includegraphics[width=0.7\columnwidth]{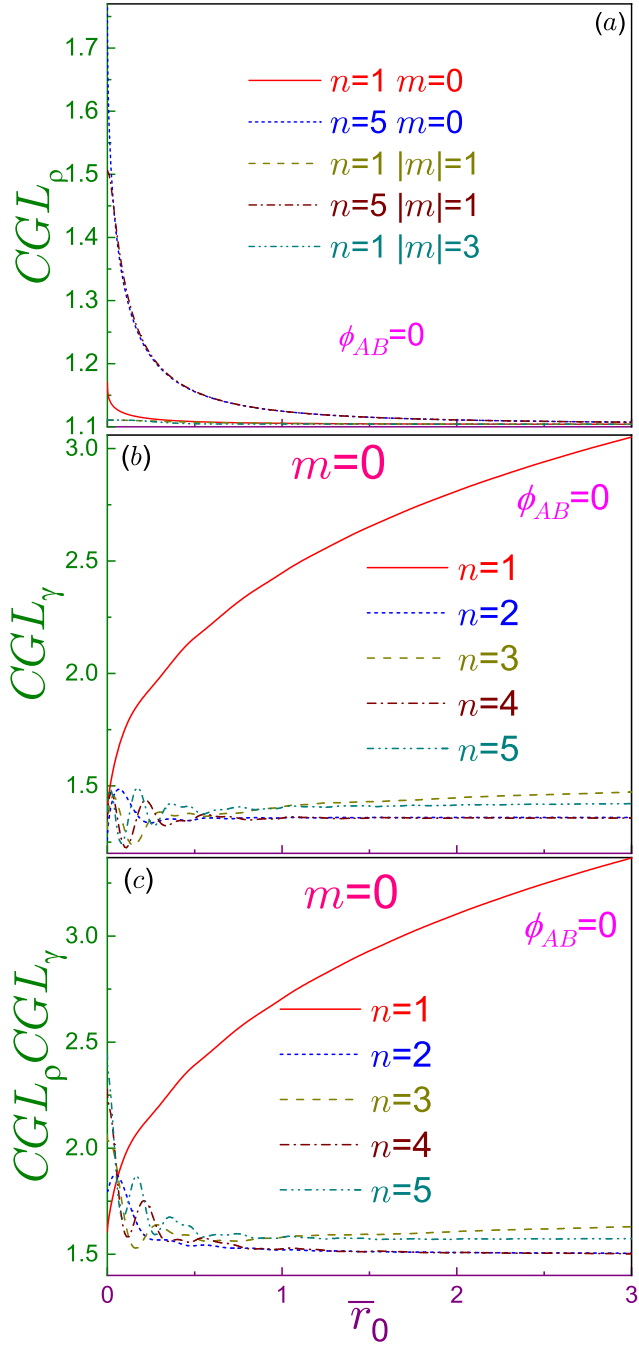}
\caption{\label{Fig_CGL1}
(a) Position $CGL_\rho$, (b) momentum $CGL_\gamma$ complexities and (c) their product in terms of the normalized inner radius $\overline{r}_0$ for several quantum numbers $n$ and $m$ denoted in each subplot.}
\end{figure}

Complexities $CGL$ are shown in Fig.~\ref{Fig_CGL1}. Their behavior is determined by the interplay between the Shannon entropy and disequilibrium. It is seen from panel (a) that position components monotonically decrease with the elongating radius. For the very thin rings, it immediately follows from Eqs.~\eqref{ShannonPositionAsymptote1} and \eqref{OnicescuPositionAsymptote1} that their asymptote is a level-independent 1D value \cite{Olendski5}:
\begin{equation}\label{CGLPositionAsymptote1}
CGL_{\rho_{nm}}\rightarrow\frac{3}{e}=1.1036\ldots,\quad\overline{r}_0\rightarrow\infty.
\end{equation}

As described in Sec.~\ref{Subsec_ShannonABfree}, at the small $\overline{r}_0$, momentum Shannon entropy, depending on the level, might exhibit irregular structure of quite chaotic oscillations and with the growth of the radius it approaches constant value. Its interaction with the corresponding Onicescu counterpart discussed above in this section leads to the patterns depicted in panel (b). Ground-state momentum complexity is a monotonically increasing function of the radius whereas the other orbitals experience at small and moderate $\overline{r}_0$ the swings contributed by Shannon multiplier and at the very thin rings they tend to the orbital-dependent limit. Essentially, the same shape is repeated by the product of the position and momentum complexities,  as exemplified by window (c).

\subsubsection{R\'{e}nyi entropy}\label{Subsec_RenyiABfree}
\begin{figure}
\centering
\includegraphics[width=0.7\columnwidth]{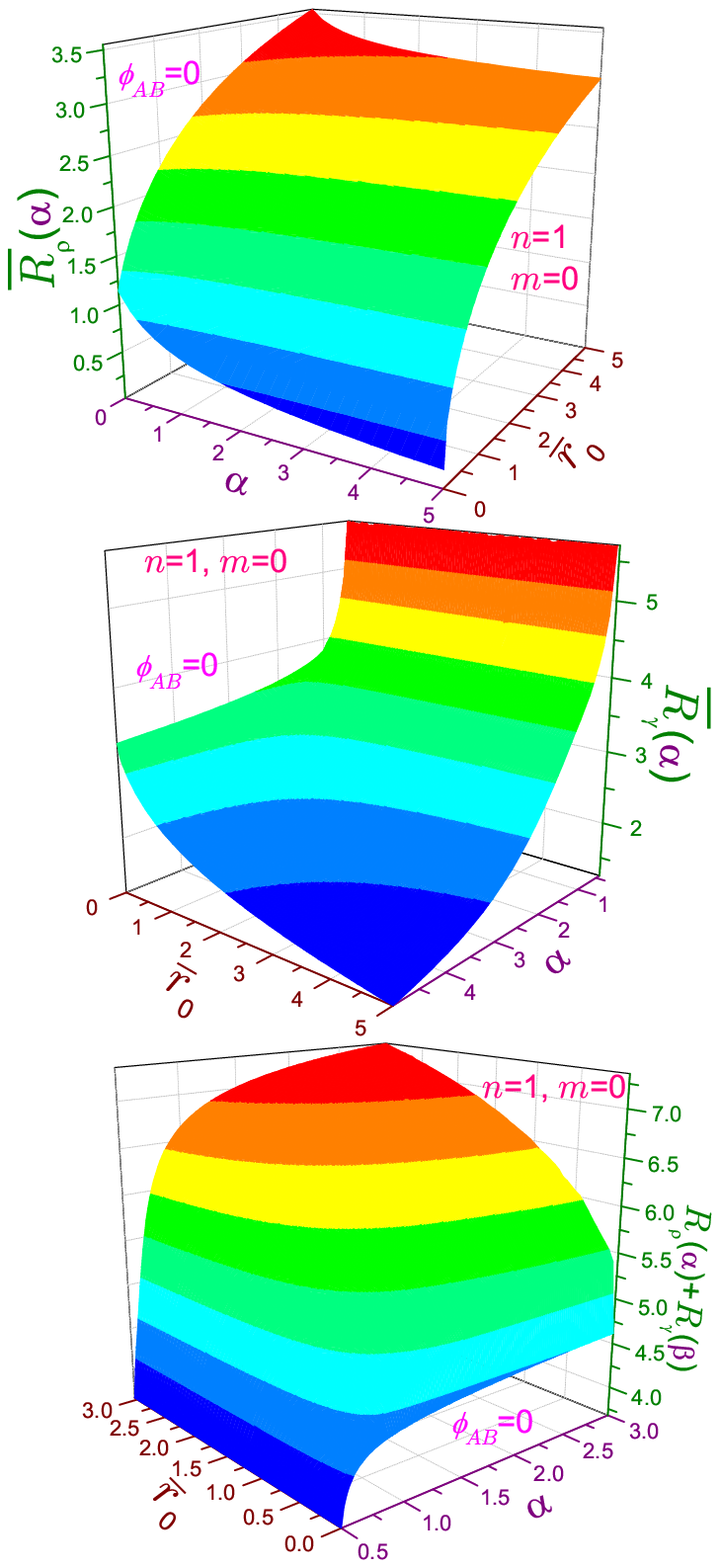}
\caption{\label{Fig_Renyi1}
AB-free ground state R\'{e}nyi entropy in terms of parameter $\alpha$ and dimensionless inner radius $\overline{r}_0$ with upper window depicting position component $\overline{R}_{\rho_{10}}(\alpha)$ and middle panel exhibiting momentum counterpart $\overline{R}_{\gamma_{10}}(\alpha)$ whereas lower subplot presenting left-hand side of the uncertainty relation, Eq.~\eqref{RenyiUncertainty1}.}
\end{figure}
At $\alpha=0$, the integral in the functional from Eq.~\eqref{Renyi1_R} yields for our geometry the area of the annulus, Eq.~\eqref{Area1}, and, accordingly, R\'{e}nyi position entropy is the same for all states:
\begin{equation}\label{RenyiPositionZero1}
\overline{R}_{\rho_{nm}}(0)=\ln\pi(2\overline{r}_0+1).
\end{equation}
In the opposite regime of the infinite parameter, one needs to use the relation:
\begin{equation}\label{RenyiPositionInfinity1}
R_{\rho,\gamma}(\infty)=-\ln\!\left(\!\!\begin{array}{c}
\rho_{max}\\
\gamma_{max}
\end{array}\!\!\right).
\end{equation}
Then, for the thin ring, it follows from the form of the radial dependence, Eq.~\eqref{Asymptote2}:
\begin{equation}\label{RenyiPositionInfinity2}
\overline{R}_{\rho_{nm}}(\infty)=\ln\pi\overline{r}_0,\quad\overline{r}_0\rightarrow\infty,
\end{equation}
which is level-independent again. Comparison of Eqs.~\eqref{RenyiPositionZero1} and \eqref{RenyiPositionInfinity2} reveals that in this regime the range of change of the position R\'{e}nyi entropy is limited by $\ln2$.

Upper panel of Fig.~\ref{Fig_Renyi1} shows evolution of the ground-state measure $\overline{R}_{\rho_{10}}$ in terms of the coefficient $\alpha$ and radius $\overline{r}_0$. At each fixed $\overline{r}_0$, R\'{e}nyi entropy is a monotonically decreasing function of its parameter what is its general property. The range in which the position functional varies shrinks at the increasing inner radius. Qualitatively, very similar dependencies are observed for all other orbitals too.

First and the most important thing to notice about the momentum entropy $R_\gamma(\alpha)$ is the fact that the lowest orbital-independent value $\alpha_{TH}$ of the semi infinite range of the R\'{e}nyi coefficient inside which this component exists is not influenced by the radius $r_0$ and is equal to its disc counterpart of two fifths \cite{Olendski1}:
\begin{equation}\label{Threshold1}
\alpha_{TH}=\frac{2}{5}.
\end{equation}
This is a direct consequence of the form of the wave vector function, Eq.~\eqref{MomentumRadial2}, and a comparison convergence test applied to the improper integral from Eq.~\eqref{Renyi1_K}. Thus, the change of the topology of the 2D Dirichlet structure from the dot to the ring does not affect the domain of the R\'{e}nyi parameter where the corresponding functional is defined. As $\alpha$ approaches from the right the threshold from Eq.~\eqref{Threshold1}, the corresponding measure logarithmically diverges, what is a general property of the momentum entropy \cite{Olendski1,Olendski3,Olendski6,Olendski7,Olendski8}. In addition, as the middle panel of Fig.~\ref{Fig_Renyi1} shows, at any fixed coefficient $\alpha$ the value of $\overline{R}_\gamma(\alpha)$ monotonically decreases with increasing the inner radius $\overline{r}_0$. At this point, it is instructive to mention that for the Volcano potential the threshold $\alpha_{TH}$, besides being controlled by the index $m$ and the AB strength, depends also on the radius \cite{Olendski3}.

Due to the conjugation from Eq.~\eqref{RenyiUncertainty2} and positiveness of the coefficients $\alpha$ and $\beta$ entering it, the corresponding uncertainty relation, Eq.~\eqref{RenyiUncertainty1}, when expressed in terms of, e.g., the R\'{e}nyi parameter $\alpha$ is \textit{not} defined if the latter is smaller than $1/2$. The upper limit of its validity is determined by the threshold from Eq.~\eqref{Threshold1} and since the Dirichlet $\alpha_{TH}$ does not exceed one half, the sum $R_{\rho_{nm}}(\alpha)+R_{\gamma_{nm}}\!\!\left(\frac{\alpha}{2\alpha-1}\right)$ is calculated in the range $[1/2,+\infty)$, as, obviously, was the QD case \cite{Olendski1} too. Lower window of Fig.~\ref{Fig_Renyi1} depicts the left-hand side of the R\'{e}nyi uncertainty relation, Eq.~\eqref{RenyiUncertainty1}, for the smallest indices $n$ and $|m|$. It is known \cite{Olendski1,Olendski3,Olendski7,Olendski8} that for the lowest-energy level the R\'{e}nyi uncertainty relation saturates at $\alpha=1/2$, what at any radius is also seen from the figure. For the excited orbitals, the left-hand side of Eq.~\eqref{RenyiUncertainty1} is always greater than its right-hand side counterpart. Explanation of these two closely related phenomenona is very similar to the geometries considered before \cite{Olendski1,Olendski3,Olendski7,Olendski8} and, for brevity, is not discussed here. It is seen from Fig.~\ref{Fig_Renyi1} that at the arbitrary $\alpha>1/2$ the sum $R_{\rho_{10}}(\alpha)+R_{\gamma_{10}}\!\!\left(\frac{\alpha}{2\alpha-1}\right)$ monotonically grows with the radius simultaneously staying an increasing function of $\alpha$; in particular, the limiting finite value at $\alpha\rightarrow\infty$ that for the QD was $4.665\ldots$ \cite{Olendski1} is pushed upwards as the ring becomes thinner.

\begin{figure}
\centering
\includegraphics[width=\columnwidth]{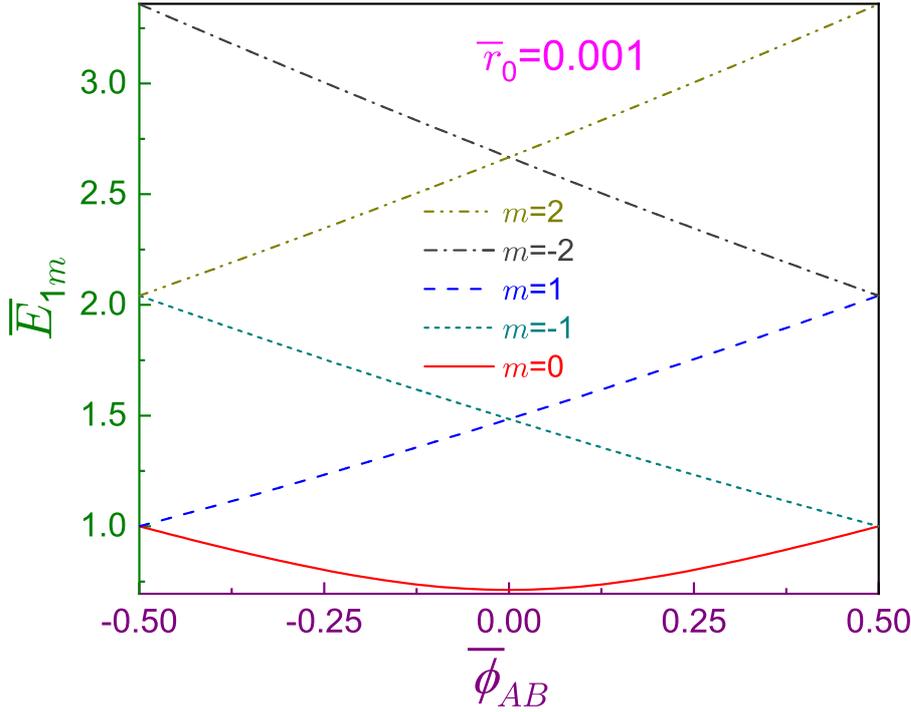}
\caption{\label{Fig_EnergiesAB1}
Ground-band, $n=1$, energy spectrum variation with the normalized AB flux $\overline{\phi}_{AB}$ for the three lowest $|m|$ specified in the legend. Dimensionless inner radius is $\overline{r}_0=0.001$.}
\end{figure}

\subsection{AB measures}\label{Subsec_ABmeasures}
\begin{figure}
\centering
\includegraphics[width=0.7\columnwidth]{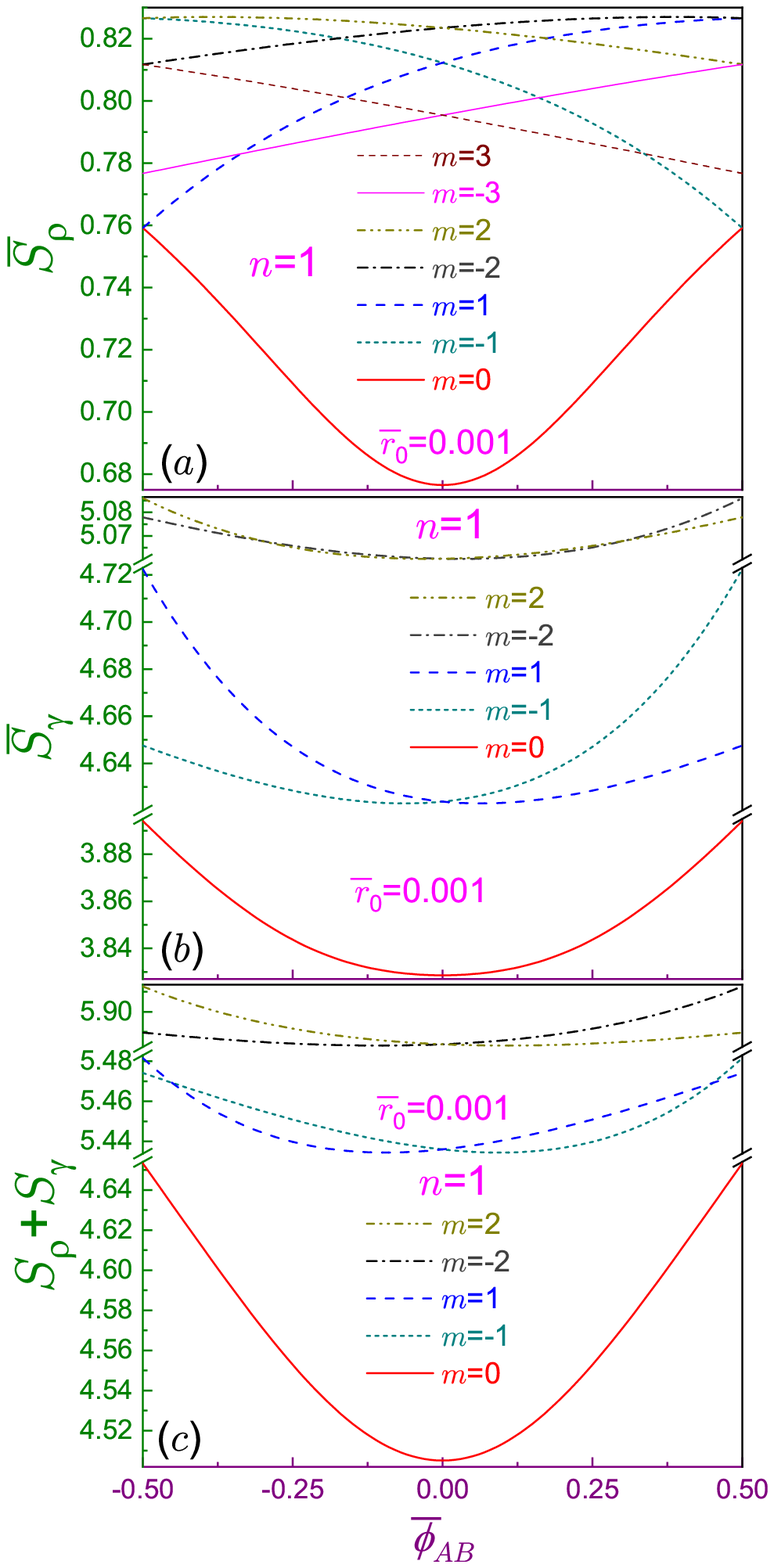}
\caption{\label{Fig_ShannonAB1}
Ground-band, $n=1$, Shannon (a) position $\overline{S}_\rho$, (b) momentum $\overline{S}_\gamma$ quantum information entropies and (c) their sum $S_\rho+S_\gamma$ versus normalized AB flux $\overline{\phi}_{AB}$ for several lowest $|m|$ specified in each legend. Dimensionless inner radius is $\overline{r}_0=0.001$. Note vertical line breaks from 3.895 to 4.62 and from 4.723 to 5.06 in panel (b) and from 4.654 to 5.434 and from 5.483 to 5.8825 in panel (c) and different scales in each split range. Dimensionless inner radius is $\overline{r}_0=0.001$.}
\end{figure}

Since the influence of the AB field is the most pronounced at the very small radius, below we will use it fixed at $\overline{r}_0=0.001$. Also, to save space, in the analysis below we do not consider R\'{e}nyi entropy.

\subsubsection{Shannon entropy}\label{Subsec_ABmeasuresShannon}
Fig.~\ref{Fig_EnergiesAB1} depicts ground-band spectrum as a function of the flux $\overline{\phi}_{AB}$. Qualitatively, the same dependencies are observed for other principal indices too. Convexity of the $m=0$ state with its minimum at $\phi_{AB}=0$ and monotonic slightly convex energy increase (decrease) for positive (negative) azimuthal quantum numbers are standard characteristics of such relationships \cite{Olendski2}. In addition, it is seen that, obviously, Eqs.~\eqref{ABrelations2} are satisfied and, in particular, for $m=0$ their explicit $\overline{r}_0$-independent values turn to the 1D Dirichlet spectrum:
\begin{equation}\label{ABrelations3}
\overline{E}_{n,0}\left(-\frac{1}{2},\overline{r}_0\right)=\overline{E}_{n,0}\left(\frac{1}{2},\overline{r}_0\right)=\overline{E}_{n,1}\left(-\frac{1}{2},\overline{r}_0\right)=\overline{E}_{n,-1}\left(\frac{1}{2},\overline{r}_0\right)=n^2,
\end{equation}
with the corresponding position wave functions degenerating to:
\begin{equation}\label{ABrelations4}
{\cal R}_{n0}\left(\pm\frac{1}{2},\overline{r}_0;r\right)={\cal R}_{n,\pm1}\left(\mp\frac{1}{2},\overline{r}_0;r\right)=\frac{1}{r^{1/2}}\,\psi_n^{(1D)}(r-r_0),
\end{equation}
what can be easily derived from the general  formulas, Eqs.\eqref{EigenValueEquation1}, \eqref{PositionRadial1} and \eqref{Normalization1}, and properties of the Bessel functions \cite{Abramowitz1}.

To understand Shannon entropy behavior of the Dirichlet ring in the AB field, one has first to clarify its dependence on the azimuthal index at the zero flux;  namely, at this small $\overline{r}_0$, similar to the QD \cite{Olendski1}, from rotationally symmetric configuration with $\overline{S}_{\rho_{1,0}}(0,10^{-3})=0.6765$, position component increases until at $|m|=2$ it reaches the maximum of $0.8235$ after which the functional monotonically decreases with growing $|m|$ (for example, at $|m|=3$ its magnitude is $0.7954$) taking negative values at quite large absolute values of the azimuthal quantum number. Diminution of the position entropy means a rise of the knowledge about the particle location which is due to the fact that it is stronger and stronger pushed against the external surface $r=r_0+a$ accumulating at it. Different situation is observed for the Volcano disc when, as mentioned above, higher $|m|$ forces the electron further and further from the origin \cite{Olendski9} causing a monotonic increase of $S_\rho$ with the magnetic index what was also the case for the energy levels \cite{Olendski2}. For this shape of the QR, this results in a  very strong similarity between the $\overline{E}-\overline{\phi}_{AB}$ and $\overline{S}_\rho-\overline{\phi}_{AB}$ characteristics\cite{Olendski2}; in particular, ground-state position entropy is a convex function of the flux and the states with positive (negative) $m$ are monotonically increasing (decreasing) convex functions of $\overline{\phi}_{AB}$ at any index. Panel (a) of Fig.~\ref{Fig_ShannonAB1} shows position measure of the Dirichlet structure versus the field. Ground-state entropy continues to exhibit a convex dependence and, similar to the Volcano shape \cite{Olendski2}, $\overline{S}_\rho$ of the $m=+1$ ($m=-1$) levels is an increasing (decreasing) function of $\overline{\phi}_{AB}$ with, however, concave shape. Mathematically, a function of one variable $f(x)$ is convex (concave) if its second derivative on the interval is positive (negative): $f''(x)>0$ [$f''(x)<0$] \cite{Boyd1}. Knowledge of this property is important in many fields of mathematics and physics, including, in particular, optimization problems \cite{Boyd1}. Even more drastic differences show up for the higher azimuthal quantum numbers, $|m|\geq2$; namely, the position entropies retain a concave form but  they have changed their growth direction: for the positive (negative) $m$, the position measures turned into the decreasing (increasing) dependencies of $\overline{\phi}_{AB}$. This feature persists for the higher azimuthal indices.  Obviously, this is due to the diminishing of the entropy at the growing number $|m|$. Let us point out that for either potential, the degeneracy of the energies expressed by Eqs.~\eqref{ABrelations2}, holds true for the position functionals $S_\rho$, $I_\rho$ and $O_\rho$ too. However, it is not the case for the momentum components of all three measures. For the Dirichlet QR, this is explained by the expression of the radial part ${\cal K}_{nm}(\overline{\phi}_{AB},\overline{r}_0;k)$ of the wave vector function $\Phi_{nm}({\bf k})$, Eq.~\eqref{MomentumRadial1}. This loss of degeneracy for $S_\gamma$ is clearly seen in window (b) of Fig.~\ref{Fig_ShannonAB1} where two vertical line breaks are used to exhibit momentum functionals for $|m|\leq2$. A comparison of the momentum Shannon entropies of the Volcano and Dirichlet annuli reveals opposite shapes of the lowest orbital, $m=0$, of each subband: concave for the former \cite{Olendski2} and convex, as panel (b) exemplifies, for the latter. In both cases, the extremum is achieved at the zero flux. Continuing this analysis, one finds that the momentum dependencies of the Dirichlet ring with $m\neq0$ are convex, $\partial^2\overline{S}_\gamma\!\!\left/\partial\overline{\phi}_{AB}^{\,2}\right.>0$, and nonmonotonic with the only minimum achieved at the nonzero  $m$-dependent field whereas for the Volcano QR the direction of change of the concave monotonic function $S_\gamma$ is opposite to its position fellow \cite{Olendski2}.

Despite the decrease of the zero-field position measure at $|m|\geq2$, the total Shannon entropy displays a monotonically rising  dependence on arbitrary $|m|$ what is explained by the same property of $S_\gamma$. A sum of the position and momentum entropies shown in subplot (c) inherits from its wave vector component a lift of the degeneracy at $\overline{\phi}_{AB}=\pm\frac{1}{2}$. Of course, for the $m=0$ orbital, $S_{\rho_{n0}}+S_{\gamma_{n0}}$ is a convex symmetric function of the flux what means that the total amount of information about electron behavior is the largest when the AB field is turned off. Interplay of the two parts makes their $m\neq0$ sum a non-monotonic convex dependence with the minimum achieved at the negative (positive) field for $m=1$ and $m\leq-2$ ($m=-1$ and $m\geq2$) whose location is slightly shifted from its momentum counterpart. For the Volcano ring, the $m\neq0$ sums are convex functions too with, however, monotonic growth (lessening) at positive (negative) $m$ \cite{Olendski2}.

\begin{figure}
\centering
\includegraphics[width=0.7\columnwidth]{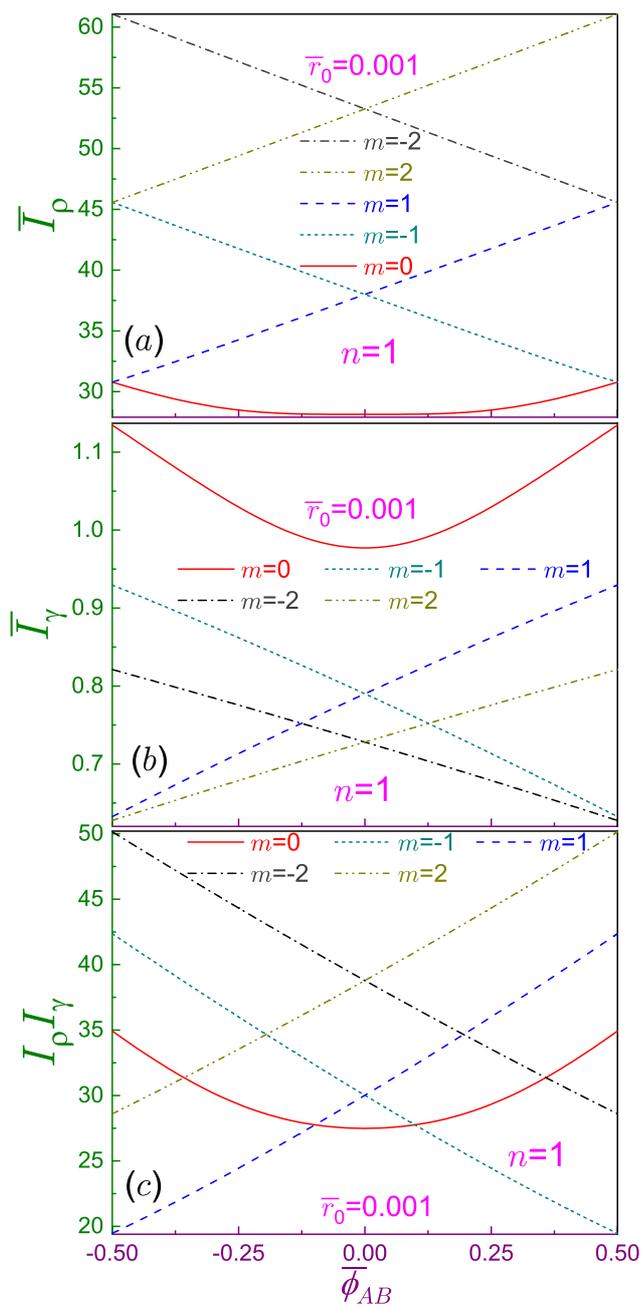}
\caption{\label{Fig_FisherAB1}
The same as in Fig.~\ref{Fig_ShannonAB1} but for the Fisher informations except that in panel (c) the products $I_\rho I_\gamma$ are shown.}
\end{figure}

\subsubsection{Fisher information}\label{Subsec_ABmeasuresFisher}
Without the field, for our chosen inner radius, in terms of the absolute value of the azimuthal quantum number, the position (momentum) Fisher information, similar to the QD\cite{Olendski1}, monotonically increases (decreases) with the product $I_\rho I_\gamma$ being a growing function of $|m|$. As was mentioned above, for the Volcano structure, the position component is independent of the flux and index $m$ \cite{Olendski2}. AB influence on the position Fisher information is displayed in Fig.~\ref{Fig_FisherAB1}(a). Unsurprisingly, a comparison with Fig.~\ref{Fig_EnergiesAB1} reveals that $\overline{I}_\rho-\overline{\phi}_{AB}$ characteristics basically mimics that of the energy spectrum; in particular, for all states $\partial^2\overline{I}_\rho\!\!\left/\partial\overline{\phi}_{AB}^{\,2}\right.>0$; for each subband $n$, the ground-orbital measure is an even function of the flux, $\overline{I}_{\rho_{n0}}\left(-\overline{\phi}_{AB};\overline{r}_0\right)=\overline{I}_{\rho_{n0}}\left(\overline{\phi}_{AB};\overline{r}_0\right)$; and, for the positive (negative) $m$, position information monotonically increases (decreases) with $\overline{\phi}_{AB}$ growing.

Applied field modifies the waveform ${\cal K}_{n0}(k)$ into the more oscillatory one. Accordingly, the Dirichlet momentum component of the lowest-energy state of each subband, similar to the Volcano ring \cite{Olendski2}, is a convex function of the field with its minimum at $\overline{\phi}_{AB}=0$, as is shown in panel (b). Fisher information of the states with positive (negative) quantum number $m$ monotonically increase (decrease) in the interval from Eq.~\eqref{Abrange1}, as was the case of the other potential too \cite{Olendski2}. The only difference between the two structures is the shape of the $m\neq0$ dependencies: for the Dirichlet annulus, $\partial^2I_\gamma\!\!\left/\partial\overline{\phi}_{AB}^{\,2}\right.<0$ whereas for the Volcano geometry it stays convex.

The most remarkable feature of the AB influence on the product of the two components is the fact that it can eliminate a monotonicity of $I_\rho I_\gamma$ on the index $m$; namely, at $\overline{\phi}_{AB}~\simeq-0.1$, decreasing from zero flux leads to the intersection of $m=0$ and $m=1$ curves, which both are convex, and then, at $\overline{\phi}_{AB}~\simeq-0.36$, the story is repeated for $m=0$ and $m=2$ lines. The same happens, as panel (c) shows, for the positive field and negative quantum numbers.

\begin{figure}
\centering
\includegraphics[width=0.7\columnwidth]{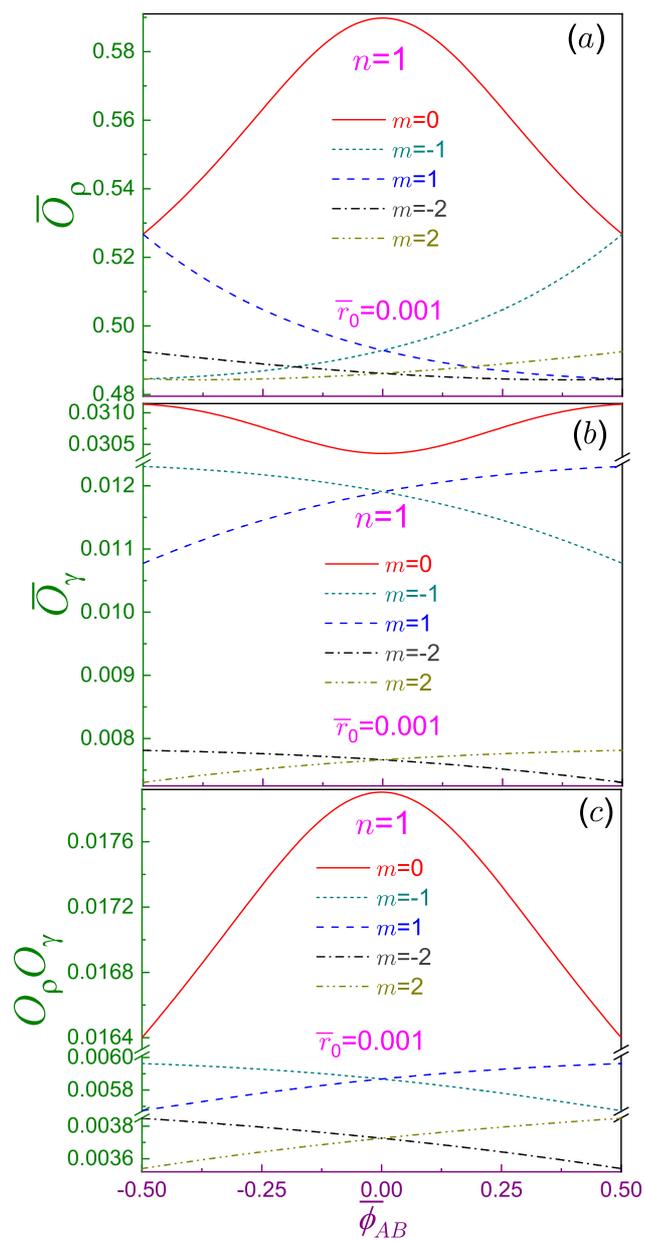}
\caption{\label{Fig_OnicescuAB1}
The same as in Fig.~\ref{Fig_FisherAB1} but for disequilibrium. Note vertical line break in subplot (b) from $0.01235$ to $0.0303$ and  two vertical breaks in panel (c) from $0.00385$ to $0.00567$ and from $0.00601$ to $0.01633$.}
\end{figure}

\subsubsection{Disequilibrium and complexity $e^SO$}\label{Subsec_ABmeasuresOnicescu}
Contrary to the Volcano ring for which both zero-field position as well as momentum Onicescu energies are monotonically decreasing functions of $|m|$ \cite{Olendski2}, for the Dirichlet annulus this is true for the wave vector component only whereas $\overline{O}_\rho$ from its $m=0$ value of $0.5898$ lessens to $|m|=2$ minimum of $0.4862$ after which it starts to grow; e.g., its $|m|=3$ magnitude is $0.5013$. Similar to the Shannon entropy, this non-monotonicity is explained by the stronger particle localization with the higher $|m|$ at the outer surface when the shape of ${\cal R}_{nm}(r)$ sharpens thus deviating further from the uniformity.

Application of the AB field transforms the ground-state position probability to the more homogeneous one resulting in a bell-like shape shown by the solid line in window (a) of Fig.~\ref{Fig_OnicescuAB1}. Alike reaction is observed for the Volcano potential too \cite{Olendski2}. The behavior of the $|m|=1$ orbitals is also very similar for both systems: positive (negative) curve decreases (increases) with the flux growing. The only difference is that for the Dirichlet ring second partial derivative $\partial^2O_\rho\!\!\left/\partial\overline{\phi}_{AB}^{\,2}\right.$ is positive whereas the Volcano QR exhibits a concave form \cite{Olendski2}. Qualitatively, for the latter geometry, the same shapes are reproduced with any other $|m|$ \cite{Olendski2}. But, due to its $\phi_{AB}=0$ non-monotonicity with respect to $|m|$, the form of the Dirichlet $\overline{O}_\rho-\overline{\phi}_{AB}$ characteristics, starting from $|m|=2$, reverses its direction retaining the convex dependence.

For both structures, the ground-state momentum disequilibrium is a convex function of the field and functionals of the orbitals with $m>0$ ($m<0$) increase (decrease) with the flux; however, for the Dirichlet QR, these dependencies are concave, as shown in panel (b), whereas for the Volcano ring, second partial derivative $\partial^2O_\gamma\!\!\left/\partial\overline{\phi}_{AB}^{\,2}\right.$ is positive.

Comparing  relative strengths of the opposite influences of the AB field on the position and momentum ground-state Onicescu functionals, one sees that, due to the their counterbalancing interaction, the result of the multiplication of the two measures is the concave function of the flux, as shown by the solid curve in subplot (c). This means that the position component dominates in the product $O_{\rho_{n0}}O_{\gamma_{n0}}$, what is true for either geometry \cite{Olendski2}. For any non-zero-index Onicescu measures, the momentum part prevails: the concave shape of $O_\rho O_\gamma$ qualitatively repeats that of $O_\gamma$.

\begin{figure}
\centering
\includegraphics[width=0.7\columnwidth]{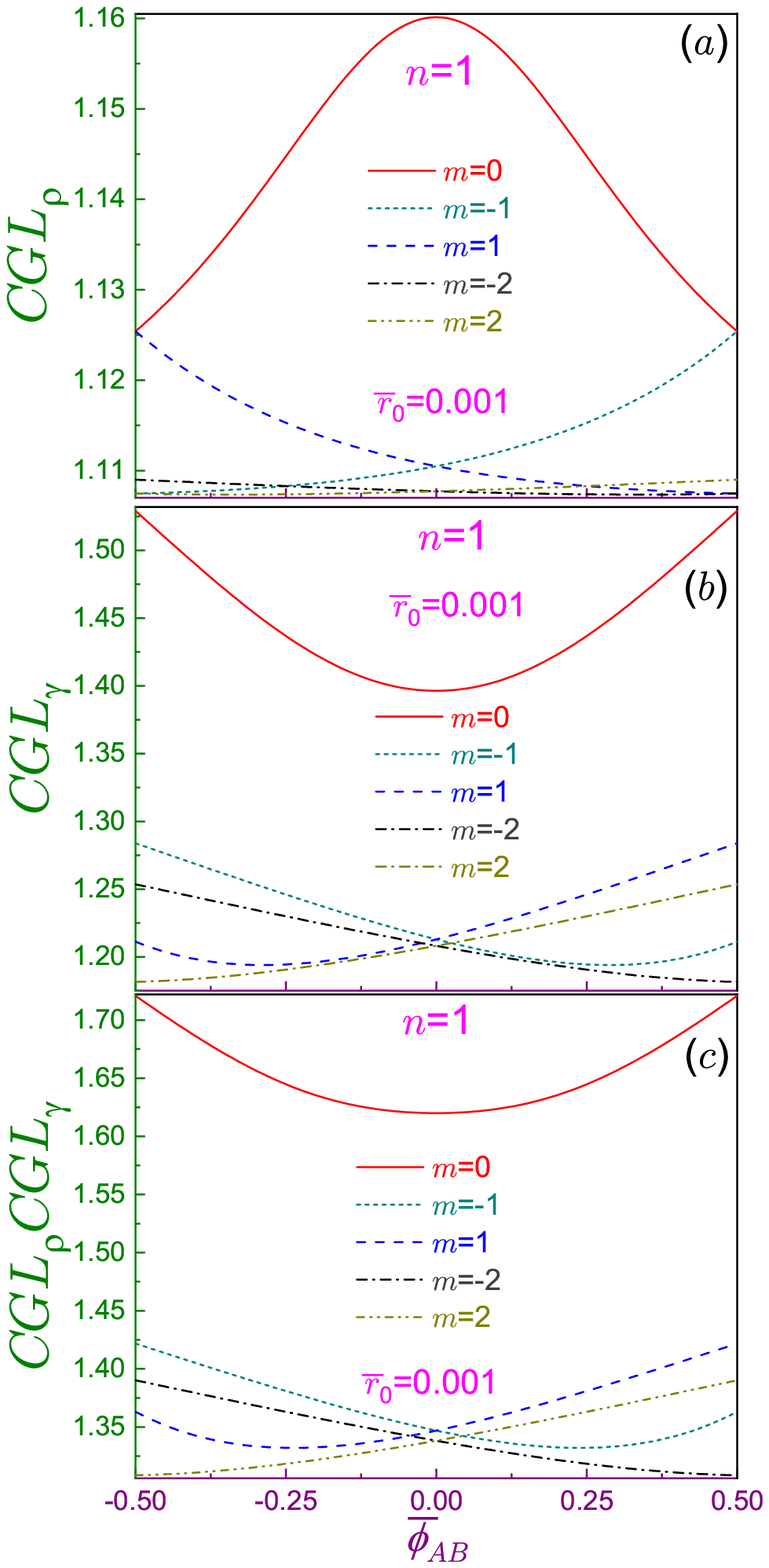}
\caption{\label{Fig_CGLAB1}
The same as in Fig.~\ref{Fig_FisherAB1} but for the complexity CGL, Eq.~\eqref{CGL1}.}
\end{figure}

Fig.~\ref{Fig_CGLAB1} displays position $CGL_\rho$ [panel (a)] and momentum $CGL_\gamma$ [subplot (b)] complexities together with their product [window (c)]. It shows that, for the ground state, the Onicescu multiplier dominates in the position component and in the product $CGL_{\rho_{n0}}CGL_{\gamma_{n0}}$: second derivatives $\frac{\partial^2}{\partial\overline{\phi}_{AB}^{\,2}}\,CGL_{\rho_{n0}}$ and $\frac{\partial^2}{\partial\overline{\phi}_{AB}^{\,2}}\left(CGL_{\rho_{n0}}CGL_{\gamma_{n0}}\right)$ carry the same sign as their Onicescu counterparts. Since momentum Shannon entropy and Onicescu energy are both convex functions of the flux, their product, of course, is shaped in the same manner. On the other hand, a comparison between solid lines in panels (a) and (b) with that from window (c) indicates a superiority of the ground-state momentum complexity over the position one. 

For $m\neq0$ orbitals, the disequilibrium exerts stronger influence on the position complexity, as compared to the Shannon entropy: qualitatively, $CGL_\rho\left(\overline{\phi}_{AB}\right)$  dependencies reproduce the Onicescu energies behavior. A more finer situation develops for the momentum product when the interplay of the convex non-monotonic Shannon part with concave monotonic Onicescu contribution yields a convex non-monotonic $CGL_\gamma-\overline{\phi}_{AB}$ characteristics with, however, different location of the minimum. This property is passed down to $CGL_\rho CGL_\gamma$ whose minimum is slightly shifted on the $\overline{\phi}_{AB}$ axis from its momentum component.

\section{Conclusions}\label{sec_Conclusions}
Quantum-information measures describe different facets of the charge distribution in position and momentum spaces. From this point of view, their knowledge is crucial for correct understanding of the processes taking place in nanostructures that are relevant for quantum-information processing. Our analysis revealed characteristic features of the Shannon and R\'{e}nyi entropies, Fisher information and disequilibrium of the hard-wall QR and their dependencies on its geometry and the AB flux. Importantly, a comparison has been made with the Volcano disc and similarities and differences between them were analyzed. The main findings include:
\begin{itemize}
\item position Shannon entropy $S_\rho$ is a monotonically increasing function of the radius and at the large $\overline{r}_0$ is proportional, regardless of the orbital, to its logarithm. This growth is explained by the increasing area of the annulus meaning a decrease of the knowledge about particle location
\item momentum Shannon component $S_\gamma$ exhibits, in general, non-monotonic dependence on the radius and for the thin rings, $r_0\gg a$, saturates to the $m$-independent value that gets larger with the principal quantum index $n$. Accordingly, a sum $S_\rho+S_\gamma$, which defines a lack of the combined information about the particle location and momentum, logarithmically increases with the radius meaning a loss of the total knowledge
\item position Fisher information $I_\rho$ essentially reproduces (with small peculiarities) energy spectrum dependence on the radius, as it should be
\item since the momentum function depicts a stronger oscillator structure with the transformation of the ring geometry from the thick to the thin one, the corresponding Fisher information $I_\gamma$, which is a measure of the wavering of the associated density, increases during this evolution
\item position disequilibrium $O_\rho$ monotonically decreases with the radius what is explained by the flattening of the wave function $\Psi({\bf r})$ whereas momentum component $O_\gamma$ behaves differently depending on the principal number: increases at the large $\overline{r}_0$ for $n=1$ and for $n>1$ saturates to the $n$-determined value
\item a change of the topology from the singly (QD, Ref.~\cite{Olendski1}) to the doubly connected domain does \textit{not} change the critical value of the R\'{e}nyi or Tsallis parameter above which the corresponding momentum measure exists. Additionally, Dirichlet threshold, Eq.~\eqref{Threshold1}, is independent of the radius
\item AB field presents an additional tool for controlling quantum-information measures. Remarkably, its influence is different for the different shapes of the confining potential of the ring; for example, for the Volcano ring, the flux $\phi_{AB}$ has no effect whatsoever on the position Fisher information \cite{Olendski2} whereas the $I_\rho-\phi_{AB}$ characteristics for the Dirichlet annulus is an almost exact replica of the energy dependence on the flux. Shape of, for example, $S_\rho(\phi_{AB})$ function might change from concave to convex if the ring geometry switches from the Dirichlet to the Volcano one. Other similarities and differences have been pointed out too.
\end{itemize}
As already mentioned in the Introduction, contemporary technologies detect and measure entanglement Shannon and R\'{e}nyi entropies of the many-body localized systems \cite{Lukin1,Niknam1,Islam1,Kaufman1,Brydges1} what expands quantum information concepts in corresponding nano devices. QR looks like one of the promising elements in quantum information processing \cite{Steiner1,Rasanen1}. Further experimental advances in this direction require deeper knowledge of its information-theoretical properties as part of the circuits with preassigned characteristics that can be easily and at will manipulated by the external influences, such as, e. g., the magnetic field. Here, a comparative analysis has been provided for the main QR quantum information measures, such as Shannon and R\'{e}nyi entropies, Fisher information and disequilibrium, with emphasis on their dependence on the ring geometry and the AB flux.
This understanding might serve as a guidance  for the optimal design of the elements for quantum information processing whose desired properties are determined by the potential used.

Above, only the Dirichlet BC was taken into account. Edge requirements strongly affect the properties of the structure \cite{Olendski1,Olendski3,Olendski5,Olendski7,Olendski8} and different BCs are used for the description of the different materials. This raises the question of investigation of the Neumann annulus with emphasis on its quantum-information measures. Also, recently, a generalization of the measures of the 2D quantum dot  has been extended to the arbitrary integer number of dimensions $\mathtt{d}\geq3$ \cite{Olendski8}; in particular, it was shown that the Dirichlet $\alpha_{TH}^D$ and Neumann $\alpha_{TH}^N$ thresholds  depend on $\mathtt{d}$ as:
\begin{subequations}\label{Ddimensional1}
\begin{align}\label{Ddimensional1_D}\alpha_{TH}^D(\mathtt{d})&=\frac{\mathtt{d}}{\mathtt{d}+3}\\
\label{Ddimensional1_N}\alpha_{TH}^N(\mathtt{d})&=\frac{\mathtt{d}}{\mathtt{d}+1}.
\end{align}
\end{subequations}
From this point of view, it would be of interest to calculate quantum-information measures of the Dirichlet and Neumann $\mathtt{d}$D cavity with inner $r_0$ and outer $r_0+a$ radii.

\begin{acknowledgements}
Research was supported by Competitive Research Project No. 2002143087 from the Research Funding Department, Vice Chancellor for Research and Graduate Studies, University of Sharjah.
\end{acknowledgements}

%
\section*{Conflict of interest}
The author  declares that he has no conflict of interest.

\section*{Data Availability}
The author declares that the data supporting the findings of this study are available within the article.


\begin{thebibliography}{}
\bibitem{Shannon1}C.E. Shannon, Bell Syst. Tech. J. {\bf 27}, 379 (1948)
\bibitem{Shannon2}C.E. Shannon, Bell Syst. Tech. J. {\bf 27}, 623 (1948)
\bibitem{Bialynicki1}I. Bia{\l}ynicki-Birula, J. Mycielski,\CMP{\bf 44}, 129 (1975)
\bibitem{Beckner1}W. Beckner,\AnM{\bf 102}, 159 (1975)
\bibitem{Coles1}P.J. Coles, M. Berta, M. Tomamichel, S. Wehner,\RMP{\bf 89}, 015002 (2017)
\bibitem{Bialynicki3}I. Bia{\l}ynicki-Birula, {\L}. Rudnicki, in: Statistical Complexity: Applications in Electronic Structure, edited by K. D. Sen (Springer, Dordrecht, 2011), chap. 1.
\bibitem{Olendski10}O. Olendski,\APB{\bf 527}, 278 (2015)
\bibitem{Olendski11}O. Olendski,\APB{\bf 528}, 865 (2016)
\bibitem{Olendski5}O. Olendski,\APB{\bf 530}, 1700324 (2018)
\bibitem{Plastino1}A.R. Plastino, A. Plastino,\PLA{\bf 181}, 446 (1993)
\bibitem{Fisher1}R.A. Fisher, Math. Proc. Cambridge Philos. Soc.  {\bf 22}, 700 (1925)
\bibitem{Frieden1}B.R. Frieden, \textit{Science from Fisher Information} (Cambridge, Cambridge, 2004)
\bibitem{Sears1}S.B. Sears, R.B. Parr, U. Dinur, Israel J. Chem. {\bf 19}, 165 (1980)
\bibitem{Plastino2}A. Plastino, G. Bellomo, A.R. Plastino, Adv. Math. Phys. {\bf 2015}, 120698 (2015)
\bibitem{Stam1}A. J. Stam, Inf. Control {\bf 2}, 101 (1959)
\bibitem{Dembo1}A. Dembo, T. M. Cover, J. A. Thomas, IEEE Trans. Inform. Theory {\bf 37}, 1501 (1991)
\bibitem{Romera2}E. Romera, P. S\'{a}nchez-Moreno, J. S. Dehesa, \CPL {\bf 414}, 468 (2005)
\bibitem{Dehesa1}J. S. Dehesa, A. Mart\'{i}nez-Finkelshtein, V. N. Sorokin, Mol. Phys. {\bf 104}, 613 (2006)
\bibitem{SannchezMoreno1}P. S\'{a}nchez-Moreno, R. Gonz\'{a}lez-F\'{e}rez, J. S. Dehesa, \NJP {\bf 8}, 330 (2006)
\bibitem{SannchezMoreno2}P. S\'{a}nchez-Moreno, A. R. Plastino, J. S. Dehesa, \jpa {\bf 44}, 065301 (2011)
\bibitem{Onicescu1}O. Onicescu,\CRA{\bf 263}, 841 (1966)
\bibitem{Ghosal1}A. Ghosal, N. Mukherjee, A. K. Roy,\APB{\bf 528}, 796 (2016)
\bibitem{Renyi1}A. R\'{e}nyi, \textit{Proceedings of the Fourth Berkeley Symposium on Mathematics, Statistics and Probability}, Berkeley University Press, Berkeley, 1961, p. 547
\bibitem{Renyi2}A. R\'{e}nyi, \textit{Probability Theory} (North-Holland, Amsterdam, 1970)
\bibitem{Bialynicki2}I. Bia{\l}ynicki-Birula,\PRA{\bf 74}, 052101 (2006)
\bibitem{Zozor1}S. Zozor, C. Vignat,\PA{\bf 375}, 499 (2007)
\bibitem{Wehner1}S. Wehner, A. Winter,\NJP{\bf 12}, 025009 (2010)
\bibitem{Jizba2}P. Jizba, J.A. Dunningham, J. Joo,\APNY{\bf 355}, 87 (2015)
\bibitem{Toscano1}F. Toscano, D. S. Tasca, {\L}. Rudnicki, S. P. Walborn, Entropy {\bf 20}, 454 (2018)
\bibitem{Hertz1}A. Hertz, N.J. Cerf,\jpa{\bf 52}, 173001 (2019)
\bibitem{Wang1}D. Wang, F. Ming, M.-L. Hu, L. Ye,\APB{\bf 531}, 1900124 (2019)
\bibitem{Tsallis1}C. Tsallis,\JSP{\bf 52}, 479 (1988)
\bibitem{Rajagopal1}A.K. Rajagopal,\PLA{\bf 205}, 32 (1995)
\bibitem{Olendski3}O. Olendski, Entropy {\bf 21},  1060 (2019)
\bibitem{Olendski1}O. Olendski,\IJQC{\bf 121},  e26455 (2021)
\bibitem{Olendski6}O. Olendski,\EJP{\bf 40}, 025402 (2019)
\bibitem{Olendski7}O. Olendski,\IJQC{\bf 120}, e26220 (2020)
\bibitem{Olendski8}O. Olendski,\EPJP{\bf 136}, 390 (2021)
\bibitem{Catalan1}R.G. Catal\'{a}n, J. Garay, R. L\'{o}pez-Ruiz,\PRE{\bf 66}, 011102 (2002)
\bibitem{Vignat1}C. Vignat, J.-F. Bercher,\PLA{\bf 312}, 27 (2003)
\bibitem{Yamano1}T. Yamano, Physica A {\bf 340}, 131 (2004)
\bibitem{Antolin1}J. Antol\'{i}n, S. L\'{o}pez-Rosa, J.C. Angulo,\CPL{\bf 474}, 233 (2009)
\bibitem{Romera1}E. Romera, \'{A}. Nagy,\PLA{\bf 372}, 6823 (2008)
\bibitem{Sen1}{\em Statistical Complexity: Applications in Electronic Structure}, ed. by  K. D. Sen (Springer, Dordrecht,   2011)
\bibitem{Toranzo1}I.V. Toranzo, P. S\'{a}nchez-Moreno, {\L}. Rudnicki, J.S. Dehesa, Entropy {\bf 19}, 16 (2017)
\bibitem{LopezRosa1}S. L\'{o}pez-Rosa, J.C. Angulo, J. Antol\'{i}n, \PA{\bf 388}, 2081 (2009)
\bibitem{Lukin1}A. Lukin, M. Rispoli, R. Schittko, M.E. Tai, A.M. Kaufman, S. Choi, V. Khemani, J. Leonard, M. Greiner, Science {\bf 364}, 256 (2019)
\bibitem{Niknam1}M. Niknam, L.F. Santos, D.G. Cory,\PRL{\bf 127}, 080401 (2021)
\bibitem{Islam1}R. Islam, R. Ma, P.M. Preiss, M.E. Tai, A. Lukin, M. Rispoli, M. Greiner, Nature (London) {\bf 528}, 77 (2015)
\bibitem{Kaufman1}A.M. Kaufman, M.E. Tai, A. Lukin, M. Rispoli, R. Schittko, P.M. Preiss, M. Greiner, Science {\bf 353}, 794 (2016)
\bibitem{Brydges1}T. Brydges, A. Elben, P. Jurcevic, B. Vermersch, C. Maier, B.P. Lanyon, P. Zoller, R. Blatt, C.F. Roos, Science {\bf 364}, 260 (2019)
\bibitem{Aharonov1}Y. Aharonov, D. Bohm,\PR{\bf 115}, 485 (1959)
\bibitem{Olendski2}O. Olendski,\PLA{\bf 383},  1110 (2019)
\bibitem{Fomin1}V.M. Fomin, \textit{Physics of Quantum Rings} (Springer, Berlin, 2014)
\bibitem{Krasnokutska1}I. Krasnokutska, J.-L.J. Tambasco, A. Peruzzo,\SR{\bf 9}, 11086 (2019)
\bibitem{Steiner1}T.J. Steiner, J.E. Castro, L. Chang, Q. Dang, W. Xie, J. Norman, J.E. Bowers, G. Moody,\PRXQ{\bf 2}, 010337 (2021)
\bibitem{Rasanen1}E. R\"{a}s\"{a}nen, A. Castro, J. Werschnik, A. Rubio, E.K.U. Gross,\PRL{\bf 98}, 157404 (2007)
\bibitem{Peshkin1}M. Peshkin,\PRe{\bf 80}, 375 (1981)
\bibitem{Skarzhinskiy1}V.D. Skarzhinskiy, Tr. Fiz. Inst. Akad. Nauk SSSR {\bf 167}, 139 (1986) [Proc. Lebedev Phys. Inst., Acad. Sci. USSR {\bf 167}, 176 (1987)]
\bibitem{Afanasev1}G.N. Afanas'ev, Fiz. Elem. Chastits At. Yadra {\bf 21}, 172 (1990) [Sov. J. Part. Nucl. {\bf 21}, 74 (1990)]
\bibitem{Avishai1}Y. Avishai, Y. Hatsugai, M. Kohmoto,\PRB{\bf 47}, 9501 (1993)
\bibitem{Olendski4}O. Olendski,\APNY{\bf 327}, 1365 (2012)
\bibitem{Buttiker1}M. B\"{u}ttiker, Y. Imry, R. Landauer,\PLA{\bf 96}, 365 (1983)
\bibitem{Srinivas1}M.D. Srinivas,\PJP{\bf 24}, 673 (1985)
\bibitem{Hall1}M.J.W. Hall,\PRA{\bf 59}, 2602 (1999)
\bibitem{Dodonov1}V.V. Dodonov and V.I. Man'ko, Tr. Fiz. Inst. Akad. Nauk SSSR {\bf 183}, 5 (1987) [Proc. Lebedev Phys. Inst., Acad. Sci. USSR {\bf 183}, 3 (1989)]
\bibitem{Matta1}C.F. Matta, M. Sichinga, P.W. Ayers,\CPL{\bf 514}, 379 (2011)
\bibitem{Flores1}N. Flores-Gallegos,\CPL{\bf 650}, 57 (2016)
\bibitem{Nascimento1}W.S. Nascimento, F.V. Prudente,\CPL{\bf 691}, 401 (2018)
\bibitem{Nascimento2}W.S. Nascimento, M.M. de Almeida, F.V. Prudente,\EJP{\bf 41}, 025405 (2020)
\bibitem{Abramowitz1}M. Abramowitz, I. A. Stegun, \textit{Handbook of Mathematical Functions} (Dover, New York, 1964)
\bibitem{McMahon1}J. McMahon,\AM{\bf 9}, 23 (1894)
\bibitem{Lowan1}A.N. Lowan, A. Hillman,\JMPC{\bf 22}, 208 (1943)
\bibitem{Kline1}M. Kline,\JMPC{\bf 27}, 37 (1948)
\bibitem{Dwight1}H.B. Dwight,\JMPC{\bf 27}, 84 (1948)
\bibitem{Waldron1}R.A. Waldron, J. Brit. Inst. Radio Eng. {\bf 17}, 577 (1957)
\bibitem{Jahnke1}E. Jahnke, F. Emde, F. L\"{o}sch, \textit{Tables of Higher Functions} (McGraw-Hill, New York, 1960)
\bibitem{Laslett1}L.J. Laslett, W. Lewish, Math. Comp. {\bf 16}, 226 (1962)
\bibitem{Gunston1}M.A.R. Gunston,\IEEETMT{\bf 11}, 93 (1963)
\bibitem{Cochran1}J.A. Cochran,\IEEETMT{\bf 11}, 546 (1963)
\bibitem{Cochran2}J.A. Cochran, J. Soc. Ind. Appl. Math. {\bf 12}, 580 (1964)
\bibitem{Willis1}D.M. Willis,\MPC{\bf 61}, 425 (1965)
\bibitem{Marcuvitz1}N. Marcuvitz, \textit{Waveguide Handbook} (Dover, New York, 1965)
\bibitem{Cochran3}J.A. Cochran,\MPC{\bf 62}, 215 (1966)
\bibitem{Bobkov1}V. Bobkov,\JMAA{\bf 472}, 1078 (2019)
\bibitem{Guo1}J. Guo, W. M\"{u}ller, W. Wang, Z. Wang,\JFA{\bf 281}, 109063 (2021)
\bibitem{Olendski9}O. Olendski, T. Barakat,\JAP{\bf 115}, 083710 (2014)
\bibitem{Landau1}L.D. Landau, E.M. Lifshitz, \textit{Quantum Mechanics (Non-relativistic Theory)} (Pergamon, New York, 1977).
\bibitem{Grochol1}M. Grochol, F. Grosse, R. Zimmermann,\PRB{\bf 74}, 115416 (2006)
\bibitem{Prudnikov1}A.P. Prudnikov, Y.A. Brychkov, O. I. Marichev, \textit{Integrals and Series} , vol. 2 (Gordon and Breach, New York, 1992)
\bibitem{Boyd1}S. Boyd, L. Vandenberghe, \textit{Convex Optimization} (Cambridge University Press, Cambridge, 2009)
\end{thebibliography}
\end{document}